\documentclass[a4paper,reqno]{amsart}
\usepackage{a4wide,color}
\usepackage{amsmath,amssymb,amsthm,mathrsfs}
\usepackage[american]{babel}
\usepackage[pdftex]{graphicx}

\newenvironment{myenumerate}{
	
	\begin{enumerate}
		\setlength{\itemsep}{0em}\setlength{\parsep}{0em}%
		\setlength{\topsep}{0em}\setlength{\parskip}{0em}%
	}
{ 	\end{enumerate} }

\numberwithin{equation}{section}

\def\R{\mathbb{R}}

\def\B{\mathcal{B}}
\def\Lebesgue{\mathcal{L}}
\def\apppf{\sharp}
\def\n{\mathbf{n}}
\def\lint{\displaystyle\int\limits}
\DeclareMathOperator{\supp}{supp}
\DeclareMathOperator{\Int}{Int}

\newtheorem{assumption}{Assumption}
\newtheorem{theorem}{Theorem}

\theoremstyle{remark}\newtheorem*{remark}{Remark}

\graphicspath{{pictures/}{./}}

\title{Pedestrian flows in bounded domains with obstacles}

\author{Benedetto Piccoli}
\address{Istituto per le Applicazioni del Calcolo ``Mauro Picone'' \\
		Consiglio Nazionale delle Ricerche \\
		Viale del Policlinico 137, 00161 Roma, Italy}
\email{b.piccoli@iac.cnr.it}

\author{Andrea Tosin}
\address{Istituto per le Applicazioni del Calcolo ``Mauro Picone'' \\
		Consiglio Nazionale delle Ricerche \\
		Viale del Policlinico 137, 00161 Roma, Italy}
\email{a.tosin@iac.cnr.it}


\begin{document}

\begin{abstract}
In this paper we systematically apply the mathematical structures by time-evolving measures developed in a previous work to the macroscopic modeling of pedestrian flows. We propose a discrete-time Eulerian model, in which the space occupancy by pedestrians is described via a sequence of Radon positive measures generated by a push-forward recursive relation. We assume that two fundamental aspects of pedestrian behavior rule the dynamics of the system: On the one hand, the will to reach specific targets, which determines the main direction of motion of the walkers; on the other hand, the tendency to avoid crowding, which introduces interactions among the individuals. The resulting model is able to reproduce several experimental evidences of pedestrian flows pointed out in the specialized literature, being at the same time much easier to handle, from both the analytical and the numerical point of view, than other models relying on nonlinear hyperbolic conservation laws. This makes it suitable to address two-dimensional applications of practical interest, chiefly the motion of pedestrians in complex domains scattered with obstacles.
\end{abstract}

\maketitle

\section{Introduction}	\label{sect:intro}
In recent years, multi-agent systems, like e.g., cell populations, bird and fish swarms, insect colonies, human crowds, have roused the interest of experts in various fields, from biology to sociology, physics, and applied mathematics. The primary reason for this scientific attention has probably to be sought in that such systems issue new stimulating challenges toward their comprehension. Indeed, they are strongly nonstandard, because of some forms of intelligence and decisional abilities, which can affect in a crucial way their evolution. In addition, they are \emph{complex systems}, in which \emph{group dynamics} plays an essential role: The \emph{collective behavior} is \emph{not} the simple superposition of individual behaviors, for complex interactions usually arise among the subjects, leading finally to completely new dynamics with respect to the case of isolated agents (cf. Krause and Ruxton \cite{KrRu}). 

Design and use of innovative analytical and numerical tools to address the study of these systems is one of the big challenges of modern applied mathematics. This is motivated, first of all, by the interest in stating a qualitative and quantitative formalization of the above-mentioned behaviors, which may help improve traditional socio-biological investigation methods by taking advantage of the typical conciseness of mathematics. On the other hand, referring in particular to human crowds, numerous engineering applications welcome the support of mathematical modeling in this nonclassical field for design and optimization purposes. For instance, in the structural engineering circle it is well-known the case of the London Millennium Footbridge, closed the very day of its opening due to macroscopic lateral oscillations of the structure developing while pedestrians crossed the bridge (cf. Dallard \emph{et al.} \cite{Millennium}). This unexpected phenomenon, caused by a synchronous lateral excitation exerted by the walkers on the footbridge, pointed out the possible relevance of crowd-structure coupling in the design of walkway infrastructures, and renewed the interest toward the investigation of these issues by means of mathematical modeling techniques (cf. Venuti \emph{et al.} \cite{MR2284944} and the main references therein). Another example is the Jamarat Bridge, located in Mina, South Arabia, $5$ km far from Mecca, reached each year by a huge crowd of pilgrims for the Hajj, the pilgrimage to Mecca that, according to the prescriptions of Islam, Muslims must carry out at least once in their lifetime. The extraordinary amount of people cramming the bridge in those occasions gave rise to serious pedestrians disasters in the nineties, with a lot of people died for overcompression. In this case, mathematical modeling and numerical simulation have already been successfully employed to study the dynamics of the flow of pilgrims, so as to highlight critical circumstances under which crowd accidents tend to occur and suggest counter-measures to improve the safety of the event (cf. Helbing \emph{et al.} \cite{HeJoAA}, Hughes \cite{Hu}).

The experimental literature on pedestrian behavior is already quite evolved. Observations and data recording by the most sophisticated techniques are nowadays extensively available \cite{HeFaMoVi,HeJo,HeJoAA,HoogTGF04,HoDa1,HoDaBo}, thanks to which numerous human walking attitudes have been pointed out. For instance, it is known that pedestrians tend to maintain a preferential direction of motion toward some targets they want to reach, but at the same time they are disposed to slightly deviate from it in order to avoid crowding. In evaluating the amount of people occupying the neighboring space, walkers take into account almost exclusively what they see ahead, because the particular position of the eyes on the head reduces their visibility field to a frontal area only. This has important consequences on the spatial configurations assumed by moving pedestrian masses: Pedestrians tend to line up in parallel lanes, unlike other kinds of agents, e.g.. birds and fishes, which instead prefer to flock. Notice that the visibility field of the latter covers also a rear area, due to the lateral positioning of their eyes on the head. Even more interesting is the well-studied case of oppositely walking pedestrian groups, in which the above-mentioned lane formation is particularly evident: When the groups cross each other, the emergence of alternate uniformly walking lanes is observed as a characteristic self-organizing tendency.

Like for many other real-world systems, collecting huge amounts of experimental data can be only a first step toward the scientific investigation of pedestrian flows. Indeed, phenomenological observations certainly enable one to gain fundamental insights into typical aspects of this complex system, as exemplified above, but they hardly allow to catch the dynamics far from equilibrium conditions. Consequently, they are mainly descriptive but scarcely predictive, as the unsteady dynamics often plays a relevant role in determining the overall evolution. Crowd motion is therefore definitely not only an engineering matter but also a challenging mathematical problem, which calls for the development of efficient modeling and simulation tools. Unlike classical physical systems dealing with inert matter, a well-coded mathematical-socio-biological theory is so far still lacking, whence the current impossibility to derive mathematical models of the behavior of the multi-agent systems discussed in this introduction from first principles. Therefore, specific models will feature necessarily phenomenological aspects. However, mathematical structures can be developed in the abstract by appealing to the most sophisticated techniques, with the aim of providing efficient modeling frameworks from both the theoretical and the applied point of view.

In this paper we are concerned with mathematical modeling of pedestrian flows. Five more sections follow this introduction, namely:
\begin{itemize}
\item Section \ref{sect:soa}, in which we account for the main literature on mathematical models of pedestrian flows currently available. Our short review covers models at both the microscopic and the macroscopic scale.
\item Section \ref{sect:mathmodmeas}, in which we illustrate the mathematical structures by time-evolving measures that we have introduced in our previous work \cite{PiTo} for the macroscopic modeling and the numerical simulation of continuous systems. We also summarize some related theoretical results, however without insisting on their proofs. The interested reader can find all details in our above-cited paper.
\item Section \ref{sect:modelpedmot}, in which we specifically apply the above framework to the modeling of pedestrian flows, discussing in particular the structure of the velocity field of pedestrians and the handling of boundary conditions. The latter issue pertains in particular to the modeling of the interactions between pedestrians and walls or obstacles.
\item Section \ref{sect:appl}, in which we address some relevant cases study by the model previously derived. We consider the motion of pedestrians in areas scattered with obstacles by simulating a group of travelers accessing one or two escalators after getting off a subway, and a group of walkers wanting to get out of a room in which two pillars partially hide the exit. Then we focus on self-organization phenomena, showing that our model is able to reproduce the typical lane formation and the emergence of alternate lanes in crossing flows. By slightly modifying a modeling detail, so as to account for a full ahead-rear visibility area of the agents, we also suggest that our model is able to qualitatively reproduce flocking phenomena, but we refrain from going too much into this issue as it is beyond the scope of the present work.
\item Section \ref{sect:conclusions}, in which we draw some conclusions and briefly sketch research perspectives.
\end{itemize}

\section{Overview of mathematical models of human crowds}	\label{sect:soa}
In this section we briefly account for the existing literature on mathematical modeling of pedestrian flows, in order to depict the scientific frame which the present paper fits in. Despite the relative novelty of the subject in the area of applied mathematics (first microscopic and macroscopic models due to Helbing, Hughes, and their respective coworkers date back to the late nineties and early two-thousands), several relevant contributions to this research line can be found already, including a number of works dealing with experimental investigation and data recording on human walking attitudes.

Pedestrians in motion within a given walking area are an essentially discrete system, in which each person plays the role of an isolated agent moving in interaction with other surrounding agents, usually with the aim of reaching a particular destination. As such, this system can be conceptually described at the microscopic scale, i.e., by a system of ordinary differential equations tracing the evolution in time of the spatial position of each single pedestrian. The possible coupling of these ODEs should reflect the interactions among pedestrians, namely the influence that each one of them has on the motion of the others. Some microscopic models of pedestrian flow are currently available in the literature, generally regarding pedestrians as rigid particles, whose motion is regulated by classical Newtonian laws of point mechanics:
$$ \dot{v}_i(t)=F_i(t), $$
where $v_i$ is the velocity of the $i$-th pedestrian, $F_i$ an overall force acting on her/him, while the index $i$ runs from $1$ to the total number $N$ of pedestrian considered in the model. The analogy with standard mechanical systems of inert matter is, however, only formal, as the force fields $F_i$ invoke nonclassical dynamic concepts resorting mainly to the ideas of:
\begin{itemize}
\item \emph{Preferred direction of motion}, i.e., the fact that pedestrians normally have a target to walk toward, which guides the main direction of their movement.
\item \emph{Comfort/discomfort} at certain distances from other pedestrians, i.e., the fact that in normal conditions pedestrians may feel uncomfortable if too close to one another and tend to look for uncrowded surrounding areas.
\end{itemize}
Therefore, in this context the $F_i$'s are not understood strictly as mechanical actions exerted by pedestrians on each other. For instance, Helbing and coworkers \cite{PhysRevE.51.4282,HeMoFaBo} introduce the concept of \emph{social} (or \emph{behavioral}) \emph{force}, which measures the internal motivation of the individuals in performing certain movements. Specifically, in their model pedestrians are regarded as points, and two main factors contribute to the definition of the social force $F_i$ acting on the $i$-th individual:
\begin{itemize}
\item A relaxation toward a \emph{desired velocity} $v_i^0$, i.e., the velocity that the $i$-th individual should possess in order to reach her/his destination as comfortably as possible:
$$ F_i^0(v_i,\,v_i^0)=\frac{v_i^0-v_i}{\tau_i}, $$
$\tau_i>0$ being the characteristic relaxation time of pedestrian $i$.
\item A \emph{repulsive effect} from either neighboring pedestrians located too close, i.e., within the so-called \emph{private sphere} of the $i$-th pedestrian:
$$ f_{ij}(r_{ij})=-\nabla{V_{ij}(r_{ij})}, \qquad j=1,\,\dots,\,N, $$
where $V_{ij}$ is a monotonically decreasing repulsive potential from pedestrian $j$ with ellipse-shaped level curves turned in the direction $r_{ij}=x_j-x_i$, or from edges of walls and obstacles found within the walking area:
$$ F_{iE}(r_{iE})=-\nabla{U_{iE}(\|r_{iE}\|)}, $$
where $U_{iE}$ is a monotonically decreasing repulsive potential from edge $E$ and $r_{iE}=x_E^i-x_i$, $x_E^i\in E$ being the edge's nearest point to pedestrian $i$.
\end{itemize}
After setting $F_i=F_i^0+\sum_{j=1}^N f_{ij}+\sum_{E}F_{iE}$, where the last sum at the right-hand side is extended to all edges $E$ of walls and obstacles that may affect the motion of the individuals, the final system of equations reads:
\begin{equation*}
	\frac{dv_i}{dt}=\frac{v_i^0-v_i}{\tau_i}-\sum_{j=1}^{N}\nabla{V_{ij}(r_{ij})}-\sum_{E}\nabla{U_{iE}(\|r_{iE}\|)},
		\qquad i=1,\,\dots,\,N.
\end{equation*}
A more refined version of the model includes, still using suitable scalar potentials, also possible attractive effects exerted on pedestrians by particular elements such as windows, displays, other people, as well as stochastic fluctuations in the behavior of each subject. The interested reader is referred to Helbing and Moln\'ar \cite{PhysRevE.51.4282}, and to the main references listed therein, for further details on these issues.

Similar behavioral considerations underlie the microscopic model of pedestrian flow by Maury and Venel \cite{MaVe}, although its formalization resorts to different mechanical guidelines. Pedestrians are still regarded as rigid particles, in particular disks of fixed radii, but the emphasis is now on a kinematic construction of their actual velocity, which does not invoke explicitly any concept of generalized force. In more detail, the preferred direction of motion is expressed by introducing a desired velocity $U$, determined by the geometry of the walking area in such a way that it drives pedestrians toward their target along the shortest path, taking into account at the same time the possible presence of intermediate obstacles. On the other hand, the repulsive effect among the individuals is meant as a geometrical constraint to avoid that disks step over one another. Specifically, it is fulfilled by first defining a set $C_q$ of admissible velocities, which at each time instant depends on the positions, denoted by $q_i$, of all pedestrians $i=1,\,\dots,\,N$, and then constructing the actual velocity as the Euclidean projection of the desired velocity of each pedestrian onto $C_q$. The evolution in time of the system is finally described as
\begin{equation*}
	q_i(t)=q_{i,0}+\lint_0^t (P_{C_q}U)(s)\,ds, \qquad i=1,\,\dots,\,N,
\end{equation*} 
where $q_{i,0}$ is the initial position of pedestrian $i$ and $P_{C_q}$ is the Euclidean projector onto $C_q$.

A further microscopic formulation of the problem is instead used by Hoogendoorn and Bovy \cite{HoBo,HoBo2}, who propose a theory of pedestrian behavior based on the concepts of walking task and walking cost. Basically, they assume that pedestrians are feedback-oriented controllers, who plan their movements on the basis of some predictions they make on the behavior of the other individuals. Predictions are dictated by a sort of cooperative or non-cooperative game theory. In either case, they are affected by a limited in time and space predictive ability of the walkers. Each pedestrian behaves so as to minimize her/his individually estimated walking cost, which is expressed by a suitable functional depending on the predicted positions of the other people.

A different approach to the description of the system uses instead partial differential equations and the theory of (possibly multidimensional) conservation laws. In this case, it is assumed that pedestrians moving within a given walking area have a continuous distribution in space, so that it makes sense to introduce their density $\rho=\rho(t,\,x)$ and to invoke some conservation principles, e.g. the conservation of mass and possibly also of linear momentum, in order to get an equation, or a system of equations, satisfied by $\rho$. One of the first threads of macroscopic models of pedestrian flow is due to Hughes \cite{Hu3,Hu,Hu2}, who proposes a two-dimensional diffusion-like equation
\begin{equation}
	\frac{\partial\rho}{\partial t}-\nabla\cdot{\left(\rho g(\rho)f^2(\rho)\nabla\phi\right)}=0
	\label{eq:Hughes}
\end{equation}
coming from the mass conservation equation supplemented by the following phenomenological closure of the velocity:
\begin{equation*}
	v=-g(\rho)f^2(\rho)\nabla\phi.
\end{equation*}
Here, $\phi$ is a scalar potential whose gradient $\nabla\phi$ determines geometrically, at each point of the domain, the main direction of motion of the individuals, whereas $f$, $g$ are scalar factors affecting the speed of pedestrians on the basis of the local crowding of the walking area, in particular considering that pedestrians try to avoid extremely high densities. By means of model \eqref{eq:Hughes} the Author studies various two-dimensional configurations of human crowd motion, including the flow of pedestrians past obstacles with application to the celebrated case of Jamarat bridge (see also Helbing \emph{et al.} \cite{HeJoAA} for an experimental investigation of this issue).

Colombo and Rosini \cite{MR2158218} introduce instead a one-dimensional macroscopic model built on a Cauchy problem for the nonlinear hyperbolic conservation law
$$ \partial_t\rho+\partial_xf(\rho)=0, \qquad x\in\R,\ t>0, $$
which reminds of the Lighthill-Whitham-Richards (LWR) model of vehicular traffic (see Lighthill and Whitham \cite{MR0072606}, Richards \cite{MR0075522}). The main difference is that the density of pedestrians exhibits two characteristic maximal values $R,\,R^\star$, with $0<R<R^\star$, at both of which the flux $f$ vanishes. In normal situations $\rho$ ranges in the interval $[0,\,R]$, where the flux is nonnegative, either strictly concave or with at most one inflection point, and has precisely one local maximum point. When the density grows above its ``standard'' maximum value, i.e., for $\rho\in (R,\,R^\star]$, it is assumed that, unlike vehicular traffic, pedestrians can still move but feel overcompressed, hence their flux is less effective than before and they enter a \emph{panic} state. In this region, the function $f$ features a trend similar to that described for $\rho\in[0,\,R]$, but with a local maximum value strictly less than the previous one. As illustrated in \cite{MR2158218,CoRo}, this allows to define a concept of solution to the above conservation law in which non-classical shocks are admitted, i.e., shocks complying with the Rankine-Hugoniot condition but possibly violating entropy criteria. As a consequence, the classical maximum principle for nonlinear hyperbolic equations, stating that the solution $\rho(t,\,x)$ remains confined within the same lower and upper bounds of the initial datum for all $x\in\R$ and all $t>0$, no longer holds true and the model is able to describe the transition of pedestrians to panic even starting from an initial density entirely bounded below the standard maximum $R$. The resulting fundamental diagram, i.e., the mapping $\rho\mapsto f(\rho)$, agrees well with experimental observations reported by Helbing \emph{et al.} in \cite{HeJoAA}.

Bellomo and Dogb\'e \cite{MR2438218}, Coscia and Canavesio \cite{MR2438214} refer instead to a two-dimensional setting, in which the walking area is represented by a bounded domain $\Omega\subset\R^2$ with possible inlet and outlet regions along the boundary $\partial\Omega$. In \cite{MR2438218} the motion of pedestrians is described by a system of two partial differential equations invoking the conservation of mass and the balance of linear momentum:
\begin{equation}
	\begin{cases}
		\partial_t\rho+\nabla\cdot(\rho v)=0 \\
		\partial_tv+(v\cdot\nabla)v=F[\rho,\,v],
	\end{cases}
	\label{eq:BellDog}
\end{equation}
where $F$ is a material model for the acceleration of the individuals depending in general in a functional way on the density $\rho$ and the velocity $v$. Conversely, in \cite{MR2438214} only the continuity equation (first equation in \eqref{eq:BellDog}) is used, and the model is made self-consistent by devising appropriate closure relations for the velocity $v$ in terms of the density $\rho$ and possibly also of its gradient $\nabla{\rho}$:
\begin{equation*}
	v[\rho,\,\nabla{\rho}](x)=\varphi[\rho,\,\nabla{\rho}]\nu(x),
\end{equation*}
where $\nu=\nu(x)$ is a unit vector identifying, at each point $x\in\Omega$, the preferred direction of motion of the crowd and $\varphi$ a scalar function expressing the speed of the latter (square brackets denote in this context functional dependence). The Authors suggest several forms for $\varphi$, referring sometimes also to the velocity diagrams of first-order vehicular traffic models.

Equations \eqref{eq:BellDog} are formally inspired by the classical fluid dynamics models of continuum mechanics, however the force (per unit mass) $F$ contains non-classical contributions accounting for:
\begin{itemize}
\item A relaxation toward a desired velocity, that makes pedestrians point in the direction of a certain target they want to reach:
\begin{equation*}
	F_1=\alpha(v_e(\rho)\nu_0-v),
\end{equation*}
where $\alpha>0$ is the inverse of the relaxation time of pedestrians, $\nu_0$ is a unit vector pointing toward the target, and finally $v_e(\rho)$ is an equilibrium speed of pedestrians depending pointwise on the crowding of the walking area.
\item A local crowding estimate, based on the pointwise values that $\nabla{\rho}$ takes along the direction of the desired velocity, which might induce pedestrians to deviate from their preferred path in order to avoid areas of high density:
\begin{equation*}
	F_2=-\frac{K^2(\rho)}{\rho}\frac{\partial\rho}{\partial\nu_0}.
\end{equation*}
\item A pressure-like term, possibly regarded as a material quantity as in the celebrated Aw-Rascle model of vehicular traffic \cite{MR1750085}, which models the reaction of pedestrians to the presence of other individuals in the surrounding environment. When also this term is included, the linear momentum balance (second equation in \eqref{eq:BellDog}) rewrites as
\begin{equation*}
	\partial_t(v+P(\rho,\,v)\nu_0)+(v\cdot\nabla)(v+P(\rho,\,v)\nu_0)=F[\rho,\,v],
\end{equation*}
where $P=P(\rho,\,v)$ is some pressure that walkers feel along the preferred path, depending on the pointwise crowding of the domain and on their current velocity.
\end{itemize}
Additional topics, like e.g. the existence of a limited visibility zone for each pedestrian when trying to evaluate the minimal crowding direction, are also discussed. In particular, special attention is paid to the characterization of the panic state and to the transition to it from regular conditions: The Authors suggest that pedestrians entering a panic state tend to follow chaotically other individuals, dropping any specific target, and therefore are mostly attracted toward areas of high density rather than seeking the less congested paths.

The usually large amount of coupled ODEs to be handled simultaneously, that microscopic models usually require, is a drawback if not from the computational point of view, thanks to the increasing power of modern calculators, certainly for analytical purposes. Among others, we recall here the difficulty to recover a global overview of the system from the knowledge of its microscopic state, possibly in connection with control and optimization issues. Macroscopic models are more suited to this, but those currently available in the literature have to face several other complications due to their intrinsic hyperbolic nature. First of all, the most natural settings for pedestrian flow problems are two-dimensional: One-dimensional models are essentially explorative, but they are unlikely to provide effective mathematical tools to deal with real applications. However, it is well known that the theory of multi-dimensional systems of nonlinear hyperbolic equations is much more complicated under both the analytical and the numerical point of view, therefore a sound mathematical mastery of such models may hardly be achieved. Secondly, the imposition of boundary conditions may be tricky in hyperbolic models, because on the one hand one is forcedly driven by the characteristic velocities in defining the inflow and the outflow portion of the boundary, while on the other hand it must be guaranteed that pedestrians do not enter or exit the domain from any point of the boundary other than the prescribed inlet and outlet regions. This issue gets even more complicated in presence of \emph{obstacles}, which have to be understood as internal boundaries to the walking area. We notice, however, that the most interesting problems for applications generally do not concern pedestrian motion in free spaces, but precisely in areas scattered with obstacles (e.g., pillars, bottlenecks, narrow passages, see Helbing \emph{et al.} \cite{HeMoFaBo}), sometimes used to force the flow of crowds in specific directions.

A time-evolving measures approach to the modeling of pedestrian flow may help overcome some of the difficulties just outlined. As we will see in much more detail in the next section, the basic idea is to use measures as mathematical tools to evaluate the degree of \emph{space occupancy} by pedestrians. In this way, it is perfectly natural to address the problem from a macroscopic, even Eulerian, point of view in spite of the intrinsic Lagrangian granularity of the system: Given a walking area $\Omega$, one describes the evolution of the system by \emph{measuring} the crowding of every subset $E\subseteq\Omega$ at each time instant. In addition, as we have shown in \cite{PiTo}, there are basically no differences in the one- or two- (and even three-) dimensional theory, therefore one can immediately tackle realistic problems without the need for conceiving preliminary one-dimensional approximations.

\section{Mathematical modeling by time-evolving measures}	\label{sect:mathmodmeas}
Canuto and coworkers have proposed in \cite{CaFaTi} an Eulerian measure-theoretical approach to the modeling of \emph{rendez-vous} problems for multi-agent systems. Inspired by their work, we have introduced in \cite{PiTo} a measure-theoretical continuous modeling framework based on conservation laws, which then has been preliminarily applied to pedestrian flow problems. The main novelty with respect to the standard literature on the subject is that such conservation laws are stated in terms of discrete-time-evolving measures rather than through hyperbolic partial differential equations, which yields several advantages over classical approaches also used by some Authors to face similar problems. From the analytical point of view, we recall: (i) An easier passage from the Lagrangian to the Eulerian description of the system, without the need for resorting heavily to regularity issues; (ii) A more straightforward establishment of the well posedness of the problem in terms of existence, uniqueness, and \emph{a priori} estimates of the solution; (iii) A direct deduction of an \emph{ad-hoc} numerical scheme for the approximate treatment of the equations, characterized by nice stability and accuracy properties as well as by a simple practical implementation. From the modeling point of view, we anticipate that this framework: (iv) Allows to address directly realistic two-dimensional applications without extra difficulties with respect to the (explorative) one-dimensional case; (v) Is particularly suited to treat nonlocal interactions among pedestrians; (vi) Gives rise to an easy and rich handling of boundary conditions as far as both wall-like and obstacle-like boundaries are concerned.

\subsection{Modeling framework}	\label{subsect:mathmodmeas-frame}
Let $\Omega\subset\R^d$, where $d=1,\,2,\,3$ from the physical point of view, be a bounded set representing the walking area of pedestrians. Topologically, we may think of it as a pathwise-connected domain possibly containing holes understood as internal obstacles to the walking area. The core of the modeling approach is the description of the space occupancy by pedestrians via a family of positive Radon measures $\{\mu_n\}_{n\geq 0}$ defined on the $\sigma$-algebra $\B(\Omega)$ of the Borel sets of $\Omega$, such that for all $E\in\B(\Omega)$ the number $\mu_n(E)\geq 0$ yields an estimate, in macroscopic average terms, of the amount of people contained in $E$ at time $n\geq 0$. Notice that this is not a pointwise information on the distribution of pedestrians in $\Omega$, but rather a hint toward their \emph{localization} within the walking area, which makes this setup conceptually meaningful in terms of a macroscopic look at the system.

Given the distribution of pedestrians in $\Omega$ at time $n$, i.e., the measure $\mu_n:\B(\Omega)\to\R_+$, the dynamics of the system toward the next time $n+1$ is described by a \emph{motion mapping} $\gamma_n:\Omega\to\Omega$, i.e., a Borel function of the form
\begin{equation}
	\gamma_n(x)=x+v_n(x)\Delta{t},
	\label{eq:motmap}
\end{equation}
where $v_n:\Omega\to\R^d$ is the velocity field of pedestrians and $\Delta{t}>0$ is the time step. In practice, $\gamma_n(x)\in\Omega$ is the position occupied at time $n+1$ by the point which at the previous time $n$ was located in $x\in\Omega$. The new distribution of pedestrians at time $n+1$ is obtained pushing the measure $\mu_n$ forward by means of the motion mapping $\gamma_n$, which corresponds to defining the new measure $\mu_{n+1}:\B(\Omega)\to\R_+$ as
\begin{equation}
	\mu_{n+1}(E)=\mu_n(\gamma_n^{-1}(E)), \qquad \forall\,E\in\B(\Omega),
	\label{eq:pushfwd}
\end{equation}
or, equivalently written, $\mu_{n+1}=\gamma_n\#\mu_n$. This equation states in formal mathematical terms the simple idea that the amount of people located in a spatial region $E$ at time $n+1$ is related to the analogous amount at the previous time $n$ along the trajectories of the motion of the pedestrians themselves. This interpretation also points out the conservation law structure of Eq. \eqref{eq:pushfwd}, which can be made even more evident by rewriting the latter in the equivalent form:
\begin{equation}
	\mu_{n+1}(E)-\mu_n(E)=-[\mu_n(\gamma_n^{-1}(E^c)\cap E)-\mu_n(\gamma_n^{-1}(E)\cap E^c)].
	\label{eq:conslaw}
\end{equation}
Considering that
\begin{equation*}
	\gamma_n^{-1}(E^c)\cap E=\{x\in E\,:\,\gamma_n(x)\not\in E\}, \qquad
	\gamma_n^{-1}(E)\cap E^c=\{x\not\in E\,:\,\gamma_n(x)\in E\},
\end{equation*}
we see that the time variation of the measure of $E$, expressed by the left-hand side of Eq. \eqref{eq:conslaw}, is related to the difference between the outgoing flux $\mu_n(\gamma_n^{-1}(E^c)\cap E)$ and the incoming flux $\mu_n(\gamma_n^{-1}(E)\cap E^c)$, which is the same conceptual idea underlying classical conservation laws in continuum mechanics.

In continuum mechanics one often deals with \emph{densities} of some continuous fields, thanks to which one can switch formally from the integral to the pointwise, namely differential, form of balance equations. In our measure-theoretical setting we can recover the concept of \emph{density of pedestrians} by assuming that at time $n$ the measure $\mu_n$ is absolutely continuous with respect to the Lebesgue measure $\Lebesgue^d$ (\emph{continuum hypothesis}), written $\mu_n\ll\Lebesgue^d$. Then Radon-Nikodym theorem implies the existence of a function $\rho_n\in L^1(\Omega)$, $\rho_n\geq 0$ a.e. in $\Omega$, such that $d\mu_n=\rho_n\,dx$, namely
$$ \mu_n(E)=\lint_E\rho_n(x)\,dx, \qquad \forall\,E\in\B(\Omega). $$
Notice that the pointwise values of $\rho_n$ in $\Omega$ are physically meaningless, as pedestrians are actually not a continuous matter. What instead makes sense is the information on the localization of pedestrians and on the crowding of the walking area provided by the measure $\mu_n$. However, having a density is useful not only from the conceptual point of view, due to the analogy with classical approaches in continuum mechanics, but also from the numerical point of view, since it is much easier to look for numerical approximations of the function $\rho_n$ rather than of the mapping $\mu_n$. The question then arises how to guarantee that, given $\mu_n\ll\Lebesgue^d$, also $\mu_{n+1}$ is absolutely continuous with respect to $\Lebesgue^d$. The answer is furnished by the following
\begin{theorem}	\label{theo:abs_cont}
Let a constant $C>0$ exist such that
\begin{equation}
	\Lebesgue^d(\gamma_n^{-1}(E))\leq C\Lebesgue^d(E), \qquad \forall\,E\in\B(\Omega).
	\label{eq:prop_gamman}
\end{equation}
If $\mu_n\ll\Lebesgue^d$ then also $\mu_{n+1}\ll\Lebesgue^d$. In addition:
\begin{myenumerate}
\item $\|\rho_{n+1}\|_1=\|\rho_n\|_1$, and
\item if $\rho_n\in L^1(\Omega)\cap L^\infty(\Omega)$ then $\rho_{n+1}\in L^1(\Omega)\cap L^\infty(\Omega)$ as well, with
$\|\rho_{n+1}\|_\infty\leq C\|\rho_n\|_\infty$.
\end{myenumerate}
\end{theorem}
For the proof of this result the interested reader is referred to \cite{PiTo}. Property \eqref{eq:prop_gamman} amounts to requiring that $\gamma_n$ does not map Lebesgue-nonnegligible subsets of $\Omega$ into Lebesgue-negligible subsets of $\Omega$; in other words, that it does not cluster the measure $\mu_n$ during the push-forward. It can be proved (see again \cite{PiTo} for the detailed calculations) that $\gamma_n$ complies with Eq. \eqref{eq:prop_gamman} if the velocity field $v_n$ is Lipschitz continuous in $\Omega$ with Lipschitz constant $L<\Delta{t}^{-1}$, which can be expressively written as:
\begin{equation*}
	\Delta{t}\vert v_n(y)-v_n(x)\vert<\vert y-x\vert, \qquad \forall\,x,\,y\in\Omega.
	\label{eq:CFL-continuous}
\end{equation*}
If all of the motion mappings $\{\gamma_n\}_{n\geq 0}$ fulfill this requirement, and moreover an initial measure $\mu_0\ll\Lebesgue^d$ is prescribed, then we deduce from Theorem \ref{theo:abs_cont} that $\mu_n\ll\Lebesgue^d$ all $n>0$ with
$$ \|\rho_n\|_1=\|\rho_0\|_1, \qquad \|\rho_n\|_\infty\leq C^n\|\rho_0\|_\infty. $$
It is worth pointing out that, owing to the above-cited Radon-Nikodym theorem, each density $\rho_n$ is unique in $L^1(\Omega)$, whence the well-posedness of Problem \eqref{eq:pushfwd} follows, along with the previous \emph{a priori} estimates, for a given initial density $\rho_0\in L^1(\Omega)\cap L^\infty(\Omega)$.

\begin{remark}[Outgoing flows]
Sometimes in applications the motion mappings $\gamma_n$ may not comply strictly with the requirement $\gamma_n(\Omega)\subseteq\Omega$, because for instance the velocity field $v_n$ points outward the domain in correspondence of some outlet portions of the boundary. For pedestrian flows, this commonly happens when exits are present along $\partial\Omega$, as we will see more specifically in Sect. \ref{sect:appl}. Nevertheless, the theory illustrated in this section still works, up to some minor technical modifications, provided one confines the attention to the restriction measures $\mu_n\llcorner\Omega$:
$$ (\mu_n\llcorner\Omega)(E):=\mu_n(E\cap\Omega), \quad \forall\,E\in\B(\R^d), $$
which essentially amounts to neglecting the dynamics possibly taking place outside $\Omega$. In particular, if $\gamma_n$ satisfies $\Lebesgue^d(\gamma_n^{-1}(E)\cap\Omega)\leq C\Lebesgue^d(E)$ for all Borel sets $E\subseteq\R^d$, then one gets again the existence and uniqueness of densities $\rho_n\in L^1(\Omega)\cap L^\infty(\Omega)$ for the measures $\mu_n\llcorner\Omega$ with respect to $\Lebesgue^d$. However, the conservation of mass $\|\rho_n\|_1=\|\rho_0\|_1$ claimed by Theorem \ref{theo:abs_cont} no longer holds, and has to be replaced by the more appropriate condition $\|\rho_n\|_1\leq\|\rho_0\|_1$.
\end{remark}

\subsection{Numerical treatment of the equations}	\label{subsect:numerics}
In order to address the spatial discretization of Eq. \eqref{eq:pushfwd} toward the design of a suitable numerical scheme for the approximate solution of the problem, we assume conveniently that $\Omega$ is the cube $[0,\,1]^d$ and partition it using a family of nested pairwise disjoint grids $\{E_i^h\}_{i=1}^{M_h}$ made of $M_h$ cubes and parameterized by the edge size $h>0$ of the latter:
\begin{equation}
	\Omega=\bigcup_{i=1}^{M_h}E_i^h, \qquad \Int{E_i^h}\cap\Int{E_j^h}=\emptyset,\quad\forall\,i\ne j.
	\label{eq:partition}
\end{equation}
In case several, say $m$, obstacles $\{O_k\}_{k=1}^{m}$, $O_k\subset\Omega$, are present within the walking area, we further suppose that each of them can be described as a suitable union of grid elements, i.e., that for each $k=1,\,\dots,\,m$ there exists a collection of indices $I_k^h\subset\{1,\,\dots,\,M_h\}$ such that $O_k=\cup_{i\in I_k^h}E_i^h$. In principle, the construction described below can be repeated by dropping any, possibly all (except properties \eqref{eq:partition}), of these simplifying assumptions, up to additional technicalities in the practical implementation of the resulting numerical scheme.

Let us introduce the measure $\lambda_h^n\ll\Lebesgue^d$ defined by the piecewise constant density
$$ P_h^n(x)=\sum_{i=1}^{M_h}\rho_{i,h}^n\chi_{E_i^h}(x), \qquad \rho_{i,h}^n\geq 0, $$
where $\chi_{E_i^h}$ is the indicator function of the set $E_i^h$, so that $d\lambda_h^n=P_h^n\,dx$. Assume in addition that an approximation $g_h^n$ of the motion mapping $\gamma_n$ is known, having the following structure:
$$ g_h^n(x)=x+\Delta{t}\sum_{i=1}^{M_h}v_n(x_i^h)\chi_{E_i^h}(x), $$
$x_i^h$ being the center of the grid cell $E_i^h$. In essence, this corresponds to an approximation of the velocity $v_n$ by the piecewise constant function
$$ u_h^n(x)=\sum_{i=1}^{M_h}v_n(x_i^h)\chi_{E_i^h}(x), $$
which makes $g_h^n$ a translation over each cell $E_i^h$, thus globally a piecewise translation over $\Omega$.

At the successive time step $n+1$ we look for a new measure $\lambda_h^{n+1}$ which is in turn absolutely continuous with respect to $\Lebesgue^d$ and piecewise constant on the partition of $\Omega$:
$$ d\lambda_h^{n+1}=P_h^{n+1}\,dx, \qquad
	P_h^{n+1}(x)=\sum_{i=1}^{M_h}\rho_{i,h}^{n+1}\chi_{E_i^h}(x). $$
Imposing that $\lambda_h^n$, $\lambda_h^{n+1}$ satisfy Eq. \eqref{eq:pushfwd} over each of the grid cells under the action of the mapping $g_h^n$, i.e.,
\begin{equation}
	\lambda_h^{n+1}(E_i^h)=(g_h^n\#\lambda_h^n)(E_i^h), \qquad \forall\,i=1,\,\dots,\,M_h,
	\label{eq:approx_pushfwd}
\end{equation}
we obtain, after some standard manipulations, the following scheme:
\begin{equation}
	\rho_{i,h}^{n+1}=\sum_{j=1}^{M_h}\rho_{j,h}^n\Lebesgue^d(E_i^h\cap g_h^n(E_j^h)), \qquad
		i=1,\,\dots,\,M_h\ \text{and\ } n\geq 0,
	\label{eq:numscheme}
\end{equation}
which relates the coefficients of $P_h^{n+1}$ to those of $P_h^n$ in a time-explicit way. Few observations are in order:
\begin{itemize}
\item It can be easily checked that the above scheme is \emph{positivity preserving}, in the sense that if $P_h^0\geq 0$ in $\Omega$ then $P_h^n\geq 0$ in $\Omega$ all $n>0$, and \emph{conservative}, that is $\|P_h^n\|_1=\|P_h^0\|_1$ all $n>0$. Moreover, under condition \eqref{eq:CFL} discussed below, which, roughly speaking, states that the magnitude of the grid size $h$ must be consistent with that of the time step $\Delta{t}$, it is also \emph{boundedness preserving}: If the initial density is discretized in such a way that $\|P_h^0\|_\infty\leq B_0$ for a certain constant $B_0\geq 0$ independent of $h$, then there exists $c\geq 0$ such that $\|P_h^n\|_\infty\leq c^nB_0$ all $n>0$ and all $h>0$.

\item The image of $E_j^h$ under $g_h^n$ is straightforwardly obtained by exploiting the piecewise translation structure of the latter as $g_h^n(E_j^h)=E_j^h+v_n(x_j^h)\Delta{t}$.

\item The measure $\lambda_h^{n+1}$ is \emph{not} the push forward of $\lambda_h^n$ by $g_h^n$, indeed Eq. \eqref{eq:approx_pushfwd} holds, by construction, on the grid cells only but in principle not for any $E\in\B(\Omega)$. It is convenient to emphasize this by introducing the notation $g_h^n\apppf\cdot$ to indicate such a sort of `approximate push forward' operated by the numerical scheme, which however coincides with the actual push forward $g_h^n\#\cdot$ when tested on the grid cells. In other words, $g_h^n\apppf\cdot$ is an operation on a piecewise constant measure which returns as output another piecewise constant measure (on the \emph{same} grid as the input measure), whose density with respect to $\Lebesgue^d$ is identified by a \eqref{eq:approx_pushfwd}-like formula. According to this notation, the scheme rewrites compactly as
$$ \lambda_h^{n+1}=g_h^n\apppf\lambda_h^n, \qquad n\geq 0. $$
\end{itemize}

The accuracy of the approximation provided by this numerical scheme is related to the fulfillment of a CFL-like condition between the time step $\Delta{t}$ and the grid size $h$, as stated by the following:
\begin{theorem}	\label{theo:stability}
Let $\gamma_n$ be invertible, smooth, with smooth inverse, and let $\Delta{t},\,h>0$ satisfy
\begin{equation}
	\Delta{t}\max_{i=1,\,\dots,\,M_h}\vert v_n(x_i^h)\vert\leq h,
	\label{eq:CFL}
\end{equation}
then:
\begin{myenumerate}
\item {(One-step stability)} There exist constants $A,\,B\geq 0$, independent of $h$, such that
$$ \sum_{i=1}^{M_h}\vert\mu_{n+1}(E_i^h)-\lambda_h^{n+1}(E_i^h)\vert\leq
	A\|\rho_n-P_h^n\|_1+Bh. $$
\item {(Multistep stability)} For each $n>0$ there exists a constant $A_n\geq 0$, independent of $h$, such that
$$ \max_{i=1,\,\dots,\,M_h}\vert\mu_n(E_i^h)-\lambda_h^n(E_i^h)\vert\leq
	\max_{i=1,\,\dots,\,M_h}\vert\mu_0(E_i^h)-\lambda_h^0(E_i^h)\vert+A_nh^d. $$
\end{myenumerate}
\end{theorem}
This stability result, proved in detail in \cite{PiTo}, allows to control
\begin{myenumerate}
\item the total variation of the localization error in one time step, and 
\item the maximum localization error after $n$ time steps
\end{myenumerate}
produced by the approximate measures $\lambda_h^n$ with respect to the actual measures $\mu_n$ in terms of the quality
\begin{myenumerate}
\item of the approximation either at the previous time step or on the initial datum, and
\item of the spatial discretization (grid size $h$).
\end{myenumerate}
We observe that hypothesis \eqref{eq:CFL} of Theorem \ref{theo:stability} amounts to controlling the maximum displacement of a grid cell, say $E_i^h$, produced by $g_h^n$, indeed for $x\in E_i^h$ we find $\vert g_h^n(x)-x\vert=\Delta{t}\vert v_n(x_i^h)\vert\leq h$. As a consequence, the number of nonempty intersections $\{E_i^h\cap g_h^n(E_j^h)\}_{j=1}^{M_h}$ in Eq. \eqref{eq:numscheme}, say $c\geq 0$, is fixed and independent of the grid size $h$ (hence also of the total number of cells $M_h$). Rather, it is determined uniquely by the dimension $d$ of the geometrical space in such a way that only grid cells $E_j^h$ sharing an edge with $E_i^h$, including obviously $E_i^h$ itself, are at most involved. The number $c$
is precisely the constant appearing in the estimate of the $\infty$-norm of $P_h^n$ mentioned above.

In the case $d=2$ this number is $9$. Using for convenience a double index notation $\{E_{ik}^h\}$ for the grid cells and all the related quantities, this means that only nine adjacent pairs of indices $(j,\,l)$ contribute at most to the intersection with the grid cell $(i,\,k)$ in Eq. \eqref{eq:numscheme}, namely
$$ (j,\,l)=(i,\,k),\ (i\pm 1,\,k),\, (i,\,k\pm 1),\, (i\pm 1,\,k\pm 1), $$
so that, denoting by $U_1$, $U_2$ the horizontal and vertical components of $v_n(x_{jl}^h)$, respectively, the coefficients $\Lebesgue^2(E_{ik}^h\cap g_h^n(E_{jl}^h))$ can be duly computed as
\begin{align*}
	\Lebesgue^2(E_{ik}^h\cap g_h^n(E_{jl}^h)) &= [U_1^+\Delta{t}\delta_{j,i-1}+
		(h-\vert U_1\vert\Delta{t})\delta_{ji}+U_1^-\Delta{t}\delta_{j,i+1}] \\
	&\phantom{=}\times [U_2^+\Delta{t}\delta_{l,k-1}+
		(h-\vert U_2\vert\Delta{t})\delta_{lk}+U_2^-\Delta{t}\delta_{l,k+1}],
\end{align*}
where $\delta_{rs}$ is Kronecker's delta and $(\cdot)^+$, $(\cdot)^-$ denote the positive and negative part of their arguments.

\section{A model for pedestrian motion}	\label{sect:modelpedmot}
As it is clear from the discussion above, the main modeling task to obtain a description of pedestrian flows from the measure-theoretical framework outlined in Subsect. \ref{subsect:mathmodmeas-frame} is the conception of a suitable velocity field $v_n$. In particular, the latter must capture some of the most relevant behavioral features of pedestrian motion, which specifically differentiate human crowds from other systems possibly describable by the same mathematical tools.

As pointed out by several Authors (see e.g., Bellomo and Dogb\'e \cite{MR2438218}, Buttazzo \emph{et al.} \cite{BuJiOu}, Colombo and Rosini \cite{MR2158218}), a preliminary important distinction has to be made on the basis of the emotional state of the crowd: One can have a flow of pedestrians in either \emph{normal} or \emph{panic conditions}, each of these states giving rise to dramatically different behavioral rules that people conform to collectively. All Authors agree neither on the same qualitative characterization of panic nor on the internal mechanisms to the crowd triggering the transition from normal conditions to panic. However, at a sufficiently general level of description, we may say that the main difference lies in the kind and the strength of \emph{interactions} that pedestrians experience with one another when entering the panic state. In our model we will not be specifically concerned with panic issues. Rather we will discuss in detail the main aspects of crowd motion in normal conditions, showing how some basic principles, together with an original modeling approach, allow to catch interesting features of the system and provide potentially powerful tools for applications in the field.

A group of pedestrians in motion within a (bounded) walking area $\Omega\subset\R^2$ usually has a \emph{target} to reach. This may be an exit, a passage, an aggregation point, that we can identify with a portion of the boundary $\partial\Omega$, possibly reduced to a point. Therefore the field $v_n$  must include a first component, call it \emph{desired velocity} $v_d$, describing the preferred direction of motion of pedestrians toward their target. When the walking area is scattered with obstacles, e.g. pillars, walls, furnishings, forbidden areas, which is the most common and interesting case in applications, the desired velocity must also account for that pedestrians may need to bypass some of them in order to get to the target.

The desired velocity gives the trajectory that each pedestrian would follow toward her/his target in the absence of other neighboring pedestrians. However, an individual walking within a group of other pedestrians is influenced in her/his motion by the presence of the latter, mainly because she/he will tend to avoid the discomfort caused by highly congested zones. Specifically, pedestrians may agree to deviate locally from their preferred path looking for uncrowded surrounding areas. Consequently they may decide to correct their desired velocity by an \emph{interaction velocity} $\nu_n$, which they determine on the basis of the crowding of the walking area in their vicinity.

The superposition of the desired velocity and the correction operated by the interaction velocity yields finally the actual velocity $v_n$ of pedestrians:
\begin{equation*}
	v_n=v_d+\nu_n.
	\label{eq:vn1}
\end{equation*}

\subsection{The desired velocity}	\label{subsect:desvel}
The desired velocity $v_d:\Omega\to\R^d$ advects each individual in the walking area toward her/his target, in the (virtual) condition of absence of other pedestrians, taking into account that she/he must bypass intermediate obstacles possibly present along the path. As such, this velocity field depends uniquely on the geometry of the domain $\Omega$ and is not affected by the distribution of pedestrians within the latter (see Maury and Venel \cite{MaVe}), $v_d=v_d(x)$.

Some basic technical assumptions we can formulate on the field $v_d$ are the following:
\begin{assumption}	\label{ass:vd}
Let the desired velocity field $v_d=v_d(x):\Omega\to\R^d$ be:
\begin{myenumerate}
\item Lipschitz continuous in $\Omega$:
$$ \exists\,T_d>0\,:\,T_d\vert v_d(y)-v_d(x)\vert\leq\vert y-x\vert, \qquad \forall\,x,\,y\in\Omega. $$
\item Uniformly bounded away from zero in $\Omega$:
$$ \exists\,V_d>0\,:\,\vert v_d(x)\vert\geq V_d, \qquad \forall\,x\in\Omega. $$
\end{myenumerate}
\end{assumption}
The assumption of Lipschitz continuity is related to the homogeneity of the movement of pedestrians in normal conditions.  The constant $T_d$, whose inverse is the Lipschitz constant of $v_d$, can be interpreted as a characteristic walking time of pedestrians. On the other hand, the uniform boundedness away from zero rules out the presence of steady desired flow in $\Omega$: Pedestrians always have a preferential direction of motion to take.

If $\Omega$ is star-shaped with respect to a point $x_0$ representing the target of pedestrians, so that every $x\in\Omega$ is connected to $x_0$ by a straight path, then a very simple form of the desired velocity is obtained as (see e.g., Coscia and Canavesio \cite{MR2438214})
$$ v_d(x)=\alpha\frac{x-x_0}{\vert x-x_0\vert}, $$
where $\alpha>0$ denotes the (possibly dimensionless) characteristic magnitude of $v_d$. We observe that such a desired velocity field complies with Assumption \ref{ass:vd} up to an arbitrarily small neighborhood of $x_0\in\Omega$.

In more complex domains the above construction of $v_d$ is no longer applicable, especially when all points $x\in\Omega$ are not connected to the target $x_0$ by a direct path due to the presence of obstacles. Then one may consider the construction of the desired velocity proposed by Maury and Venel \cite{MaVe}, which entails the identification of some intermediate targets (often coinciding with some corners of the obstacles) to be preliminarily reached until the final target $x_0$ is directly accessible. However, such a field $v_d$ is in general not as smooth as suggested by Assumption \ref{ass:vd} above, because of possible sharp swerves when approaching an obstacle. In addition, its generation gets soon laborious for an increasing number of obstacles or for variations in their mutual dispositions.

The method we want to pursue here is based on the introduction of a scalar potential $u:\Omega\to\R$, which identifies attractive and repulsive zones for pedestrians. We assume that $u$ satisfies the Laplace equation:
\begin{equation}
	\Delta{u}=0 \quad \text{in\ } \Omega
	\label{eq:Laplace}
\end{equation}
and recover the desired velocity field of pedestrians from its (normalized) gradient as
\begin{equation}
	v_d(x)=\alpha\frac{\nabla{u}(x)}{\vert\nabla{u}(x)\vert}, \qquad \alpha>0.
	\label{eq:vd}
\end{equation}
It will be occasionally useful in the sequel to refer to the unit vector $\hat{v}_d(x)=\nabla{u}(x)/\vert\nabla{u}(x)\vert$ to identify the direction of the desired velocity.

Equation \eqref{eq:Laplace} needs to be supplemented by proper boundary conditions in order to yield a unique solution $u$. In our applications we will consider essentially the two most common types of boundary conditions, namely:
\begin{myenumerate}
\item \emph{Dirichlet boundary condition}, consisting in prescribing the value of the potential $u$ along the boundary of the domain.
\item \emph{Neumann boundary condition}, which assigns instead the value of the normal derivative $\frac{\partial u}{\partial\n}$ along the boundary, $\n$ being the outward normal unit vector to $\partial\Omega$. We observe that $\frac{\partial u}{\partial\n}=\nabla{u}\cdot\n$, therefore, in view of Eq. \eqref{eq:vd}, this kind of boundary condition amounts to controlling, in a way, the normal component of $v_d$ on $\partial\Omega$.
\end{myenumerate}
According to the type of boundary under consideration, different conditions might in principle apply to different portions of $\partial\Omega$. The ultimate choice often depends on the specific geometry of the domain and on the application at hand, however, at a sufficiently general level of description, we can state the following guidelines:
\begin{itemize}
\item The outer boundary $\partial\Omega_{out}$ of $\Omega$ usually defines the walls delimiting the walking area, which may contain doors or, more in general, exits that pedestrians aim at. The latter form a subset of $\partial\Omega_{out}$ that we denote by $\Gamma_T$. To translate the will of pedestrians to reach these zones while keeping away from walls, one sets the potential $u$ to zero along the walls and to a positive value in correspondence of the targets, for instance:
\begin{equation}
	\begin{cases}
		u=0 & \text{on\ } \partial\Omega_{out}\setminus\Gamma_T \\
		u=1 & \text{on\ } \Gamma_T.
	\end{cases}
	\label{eq:bcu-out}
\end{equation}
Notice that, in order to define a desired velocity \eqref{eq:vd} leading pedestrians to the exits, the specific potential of $\Gamma_T$ is irrelevant provided it is positive, for the gradient $\nabla{u}$ will point then toward $\Gamma_T$ due simply to the growth of $u$ in that direction.

\begin{figure}[t]
\centering
\includegraphics[width=0.4\textwidth,clip]{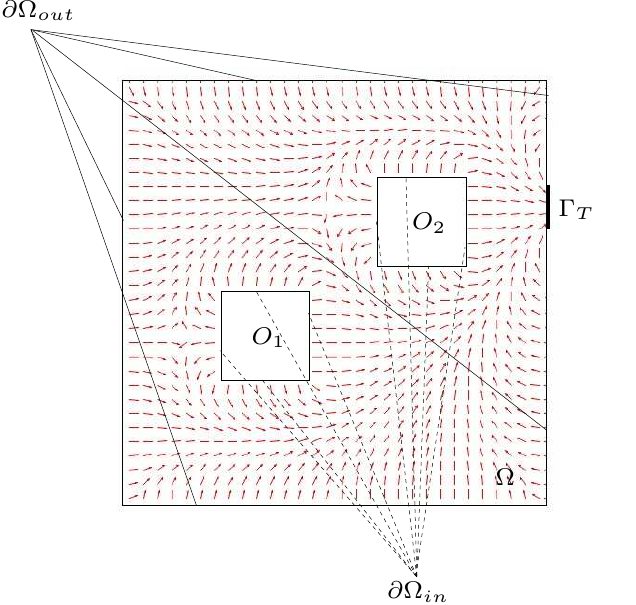}
\includegraphics[width=0.4\textwidth,clip]{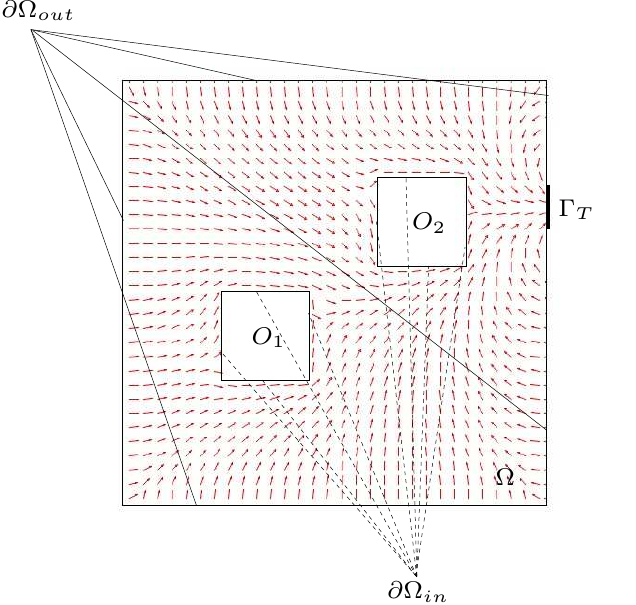}
\caption{Desired velocity field $v_d$ \eqref{eq:vd} in a domain with $m=2$ obstacles. Left: Dirichlet boundary condition  \eqref{eq:bcu-in-D} and right: Neumann boundary condition \eqref{eq:bcu-in-N} on $\partial\Omega_{in}$.}
\label{fig:bcu}
\end{figure}

\item The inner boundary $\partial\Omega_{in}$ is possibly determined by the edges of the obstacles scattered within the walking area, that pedestrians must bypass when trying to get to $\Gamma_T$. In principle, one can model their repulsive effect by prescribing a Dirichlet boundary condition of the form
\begin{equation}
	u=0 \quad \text{on\ } \partial\Omega_{in}.
	\label{eq:bcu-in-D}
\end{equation}
Coupled with condition \eqref{eq:bcu-out} above, this gives rise globally to a desired velocity field pointing outward the obstacles and the outer walls and inward the targets. This is a simple consequence of the maximum principle for the Laplace equation \eqref{eq:Laplace}, which, owing to such a set of boundary conditions, states $0\leq u\leq 1$ a.e. in $\Omega$, so that the direction of $\nabla{u}$ is from $\partial\Omega_{out}\setminus\Gamma_T\cup\partial\Omega_{in}$, where the potential attains its minimum, to $\Gamma_T$, where the potential attains instead its maximum.

It is worth noticing that the boundary condition \eqref{eq:bcu-in-D} usually produces a highly repulsive effect from the obstacles (see Fig. \ref{fig:bcu}, left), while in certain situations, and depending on the kind of obstacle, pedestrians may agree to flow by them, simply avoiding head-on collisions. This can be rendered by controlling the normal component of the desired velocity:
\begin{equation}
	\nabla{u}\cdot\n=0 \quad \text{on\ } \partial\Omega_{in}
	\label{eq:bcu-in-N}
\end{equation}
while allowing free tangential sliding of pedestrians along the walls of the obstacles (Fig. \ref{fig:bcu}, right).
\end{itemize}

We remark that the potential $u$ generated by Eq. \eqref{eq:Laplace} is smooth for domains with smooth boundary (and we can assume this is our case, up to possibly round off slightly the sharp corners of $\Omega$ and of the obstacles), hence $\nabla{u}$ is in particular Lipschitz continuous in $\Omega$. In addition, $u$ being a harmonic function, it results $\vert\nabla{u}(x)\vert\ne 0$ for a.e. $x\in\Int{\Omega}$, for $u$ can have in $\Int{\Omega}$ at most saddle points, due to topological reasons when imposing Dirichlet boundary conditions (cf. Fig. \ref{fig:bcu}, left). It follows that $\hat{v}_d$ is in turn Lipschitz continuous almost everywhere in $\Int{\Omega}$, thus finally our desired velocity $v_d$ complies almost everywhere with the requirements of Assumption \ref{ass:vd}. Notice that in case of Neumann boundary conditions, and at least for relatively simple geometries of the walking area, we do not expect instead saddle points either (cf. Fig. \ref{fig:bcu}, right). In any case, saddle points feature instable manifolds, which prevents the appearance of spurious aggregation points within the walking area.

\begin{figure}[t]
\centering
\includegraphics[width=0.18\textwidth,clip]{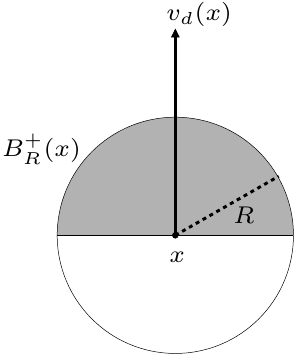}\qquad
\includegraphics[width=0.18\textwidth,clip]{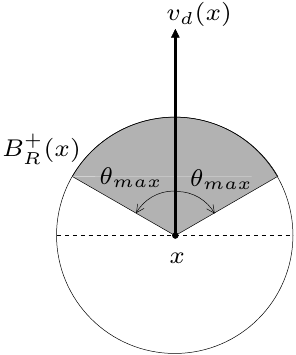}\qquad
\includegraphics[width=0.18\textwidth,clip]{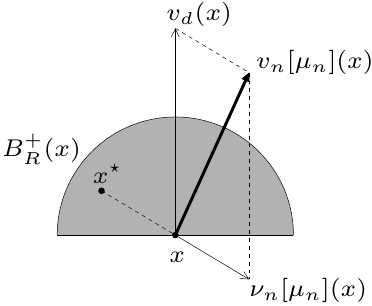}
\caption{Left and center: The interaction neighborhood $B_R^+(x)$ for a maximum angle of visibility $\theta_{max}=\pi/2$ and $0<\theta_{max}<\pi/2$, respectively. Right: Construction of the velocity $v_n[\mu_n](x)$ as the superposition of the desired velocity $v_d(x)$ and the interaction velocity $\nu_n[\mu_n](x)$. The point $x^\star$ is the center of mass of pedestrians within $B_R^+(x)$.}
\label{fig:int_neighb}
\end{figure}

\subsection{Interaction velocity}	\label{subsect:intvel}
The interaction velocity $\nu_n:\Omega\to\R^d$ expresses the fact that pedestrians are disposed to review and modify locally their preferred trajectory looking for uncrowded surrounding areas, because they feel uncomfortable when too close to one another. Unlike the desired velocity, $\nu_n$ does not only depend on the geometry of the domain but also on the current distribution of pedestrians in $\Omega$, i.e., on the measure $\mu_n$, which we emphasize by using the notation $\nu_n=\nu_n[\mu_n]$. Specifically, we assume that each pedestrian evaluates the occupancy of a certain neighborhood of her/his position $x\in\Omega$, call it the \emph{interaction} (or \emph{visibility}) \emph{neighborhood}, and then corrects the desired direction $v_d(x)$ by steering clear of the most congested zones she/he finds within that neighborhood (Fig. \ref{fig:int_neighb}).

Let $B_R(x)$ denote the $d$-dimensional ball of radius $R>0$ centered at $x\in\Omega$, then we take the interaction neighborhood of pedestrians located in $x\in\Omega$ to be (Fig. \ref{fig:int_neighb}, left)
\begin{equation*}
	B_R^+(x)=\{y\in B_R(x)\,:\,(y-x)\cdot v_d(x)\geq 0\},
	\label{eq:intneighb}
\end{equation*}
that is the half-ball in the direction of $v_d(x)$. This corresponds to the idea that pedestrians behave mainly like anisotropic particles, in the sense that they look ahead but not behind within a certain distance $R$ from themselves. By allowing $\hat{r}(x,\,y)\cdot\hat{v}_d(x)\geq\cos{\theta_{max}}$ in the definition above, where $\hat{r}(x,\,y)$ is the unit vector in the direction of $y-x$ and $\theta_{max}\in(0,\,\pi/2]$, one gets as interaction neighborhood the angular sector of $B_R^+(x)$ depicted in Fig. \ref{fig:int_neighb} center, with $\theta_{max}$ representing the maximum visibility angle of pedestrians (cf. Banos and Charpentier \cite{BaCh}, Bellomo and Dogb\'e \cite{MR2438218}).

We model the tendency of pedestrians to avoid congested areas by first detecting the center of mass $x^\star$ of the neighborhood $B_R^+(x)$:
$$ x^\star=\frac{1}{\mu_n(B_R^+(x))}\lint_{B_R^+(x)}y\,d\mu_n(y) $$
and then assuming that the correction $\nu_n[\mu_n]$ to $v_d$ amounts to a removal from $x^\star$, interpreted as an indicator of the location of average maximum crowding (Fig. \ref{fig:int_neighb}, right). We set then:
\begin{equation}
	\nu_n[\mu_n](x)=p_\nu[\mu_n](x)(x-x^\star)=
		\frac{p_\nu[\mu_n](x)}{\mu_n(B_R^+(x))}\lint_{B_R^+(x)}(x-y)\,d\mu_n(y),
	\label{eq:nun}
\end{equation}
where $p_\nu[\mu_n]:\Omega\to\R_+$ is related to the strength of the interaction among pedestrians. Some possible choices of this mapping are:
\begin{myenumerate}
\item $p_\nu[\mu_n](x)=\beta\mu_n(B_R^+(x))/R$ for $\beta>0$ (conveniently understood as dimensionless), which corresponds to a mass-dependent interaction (the more the people the stronger the interaction).
\item $p_\nu[\mu_n]=\beta/R$ for again $\beta>0$, which gives a mass-independent interaction (few people count as much as many people, the maximum strength of the interaction being fixed by $\beta$, indeed in this case $\vert\nu_n[\mu_n](x)\vert\leq\beta$ all $x\in\Omega$).
\end{myenumerate}

\begin{figure}[t]
\centering
\includegraphics[width=0.4\textwidth,clip]{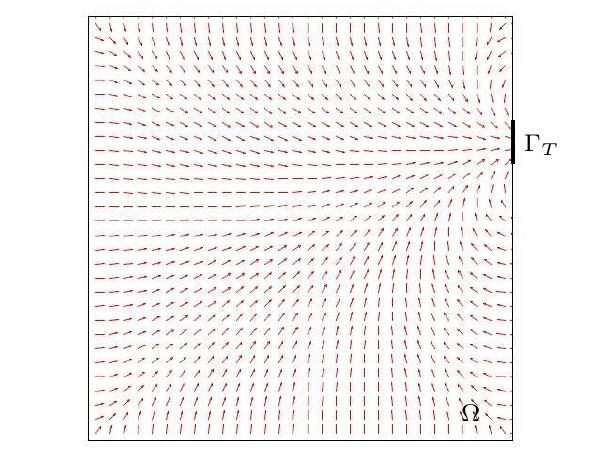}
\includegraphics[width=0.4\textwidth,clip]{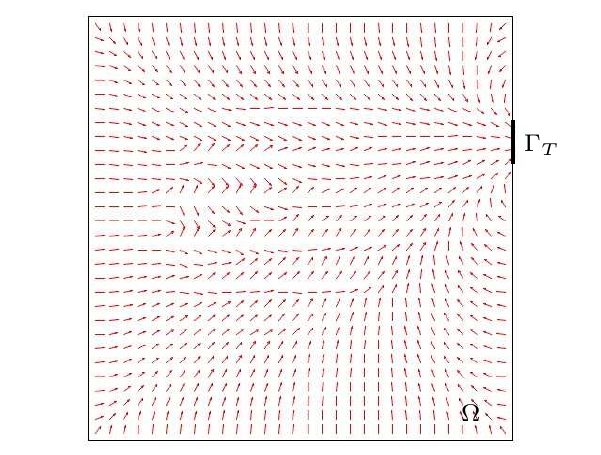}
\caption{Correction of a desired velocity field $v_d$ (left) by the superposition of the interaction velocity $\nu_n[\mu_n]$ to give the final velocity $v_n$ (right).}
\label{fig:velocity}
\end{figure}

Using the regularity hypotheses on $v_d$ expressed by Assumption \ref{ass:vd}, it is possible to prove that option (i) yields a Lipschitz continuous interaction velocity field $\nu_n[\mu_n]$ (the same is not true for option (ii), as Lipschitz continuity of $\nu_n[\mu_n]$ is broken by the presence at the denominator of $\mu_n(B_R^+(x))$, which can be arbitrarily close to zero for particular choices of the distribution of pedestrians). In short, one first proves that for $x_1,\,x_2\in\Omega$, and under the inductive assumption that $\mu_n$ is bounded from above by $\Lebesgue^d$ (which is, for instance, the case $\mu_n\ll\Lebesgue^d$ with $\rho_n\in L^1(\Omega)\cap L^\infty(\Omega)$ discussed in Sect. \ref{sect:mathmodmeas}), the estimate of $\vert\nu_n[\mu_n](x_2)-\nu_n[\mu_n](x_1)\vert$ can be essentially reduced to the estimate of $\Lebesgue^d(B_R^+(x_1)\triangle B_R^+(x_2))$. Then one shows that for $x_1,\,x_2$ sufficiently close the vectors $v_d(x_1)$, $v_d(x_2)$ point in like directions, that is $v_d(x_1)\cdot v_d(x_2)>0$, which, roughly speaking, means that the neighborhoods $B_R^+(x_1)$, $B_R^+(x_2)$ almost coincide. This allows to control the Lebesgue measure of the non-overlapping parts in terms of the distance between $x_1$ and $x_2$ as
$$ \Lebesgue^d(B_R^+(x_1)\triangle B_R^+(x_2))\leq K\vert x_2-x_1\vert $$
for a constant $K>0$ independent of $x_1,\,x_2$, which finally yields the desired Lipschitz estimate on $\nu_n[\mu_n]$. Analogous calculations are developed in detail in \cite{PiTo} for an interaction velocity built on the simpler interaction neighborhood $B_R(x)$.

Notice that the interaction velocity \eqref{eq:nun} gives rise to a nonlinear closure for the velocity $v_n$ of pedestrians, which in turn depends on $\mu_n$ (cf. Fig. \ref{fig:velocity}):
\begin{equation}
	v_n[\mu_n](x)=v_d(x)+\nu_n[\mu_n](x).
	\label{eq:vn2}
\end{equation}
If, for a given $n>0$, $\mu_n\ll\Lebesgue^d$ and $\nu_n[\mu_n]$ is Lipschitz continuous, then we can apply the theory developed in \cite{PiTo} for nonlinear fluxes $\mu_nv_n[\mu_n]$ and recover inductively the well-posedness of Problem \eqref{eq:pushfwd} expressed by Theorem \ref{theo:abs_cont} in terms of existence and uniqueness of the densities $\rho_n\in L^1(\Omega)\cap L^\infty(\Omega)$, along with \emph{a priori} estimates on their $L^1$- and $L^\infty$-norms.

\begin{remark}[Correction to $\nu_n$ near the boundaries of the domain]
Equation \eqref{eq:nun} implicitly considers in the integration only the portion of $B_R^+(x)$ contained in $\Omega$, i.e., $B_R^+(x)\cap\Omega$, because $\supp{\mu_n}\subset\Omega$ entails that obstacles and perimeter walls are $\mu_n$-negligible sets. However, it can be questioned that $\mu_n\equiv 0$ on $\Omega^c$ leads erroneously to treat obstacles and walls as uncongested areas, giving then rise to an interaction term $\nu_n[\mu_n](x)$ pointing strongly in their direction, especially for points $x\in\Omega$ sufficiently close to $\partial\Omega$ (namely, such that $B_R^+(x)\cap\Omega^c\ne\emptyset$). In some situations, this may counter the effect of the desired velocity $v_d(x)$ to such an extent that, when summing to obtain $v_n[\mu_n](x)$, one gets a velocity field pointing outside the domain. To correct this drawback it is sufficient to compute the interaction velocity with respect to a new measure $\hat{\mu}_n$ such that:
\begin{equation*}
	d\hat{\mu}_n=
		\begin{cases}
			d\mu_n & \text{in\ } \Omega \\
			Mdx & \text{in\ } \Omega^c,
		\end{cases}
\end{equation*}
which extends $\mu_n$ outside $\Omega$ by prescribing an \emph{equivalent} (or \emph{effective}) \emph{density} to obstacles and walls. The parameter $M\geq 0$ can be tuned on the basis of the discomfort produced on the crowd by the vicinity to the walls, and may possibly vary from wall to wall. Notice that $\hat{\mu}_n$ is absolutely continuous with respect to $\Lebesgue^d$ whenever $\mu_n$ is, with density $\hat{\rho}_n=\rho_n\chi_{\Omega}+M\chi_{\Omega^c}$.

The expression of $\nu_n$ modifies then in
\begin{equation}
	\nu_n[\mu_n](x)=\frac{p_\nu[\hat{\mu}_n](x)}{\hat{\mu}_n(B_R^+(x))}
		\lint_{B_R^+(x)}(x-y)\,d\hat{\mu}_n(y),
	\label{eq:nun_imprvd}
\end{equation}
which, from the modeling point of view, renders the fact that obstacles and walls contribute to the discomfort felt by pedestrians, in such a way that people might prefer to stay close to one another rather than being compressed against the walls.
\end{remark}

\begin{remark}[Retrograde flows]
The interaction velocity $\nu_n[\mu_n]$ is, by construction, opposite to the desired velocity $v_d$. It may therefore happen that the resultant $v_n[\mu_n]$ is in turn opposite to $v_d$, which gives rise to pedestrians locally (in space and time) walking away their targets (\emph{retrograde flow}). If the choice $p_\nu[\mu_n]=\beta/R$ is made, it is possible to rule this possibility out by controlling the relative magnitude of the parameters $\alpha,\,\beta$ under the requirement $v_n[\mu_n](x)\cdot\hat{v}_d(x)\geq 0$ all $x\in\Omega$, which implies:
$$ \alpha+\nu_n[\mu_n](x)\cdot\hat{v}_d(x)\geq 0. $$
Since Cauchy-Schwartz inequality implies $\vert\nu_n[\mu_n](x)\cdot\hat{v}_d(x)\vert\leq\vert\nu_n[\mu_n](x)\vert\leq\beta$, we have that the above relation certainly holds true if the following more restrictive condition is satisfied:
$$ \alpha-\beta\geq 0, $$
whence the simple criterion stating that the magnitude of $v_d$ must be greater than the maximum possible of $\nu_n[\mu_n]$.
\end{remark}

\begin{figure}[t]
\centering
\begin{minipage}[c]{0.49\textwidth}
	\centering
	\includegraphics[width=0.49\textwidth,clip]{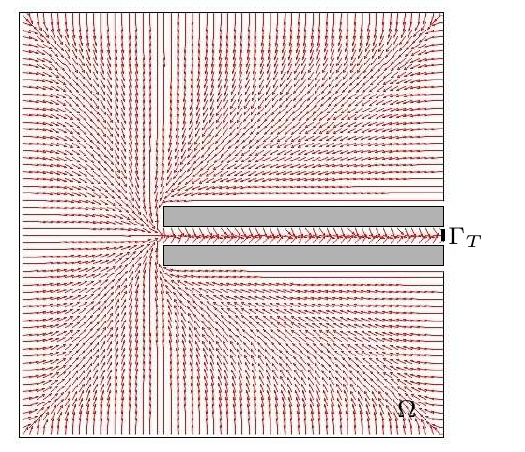}
	\includegraphics[width=0.49\textwidth,clip]{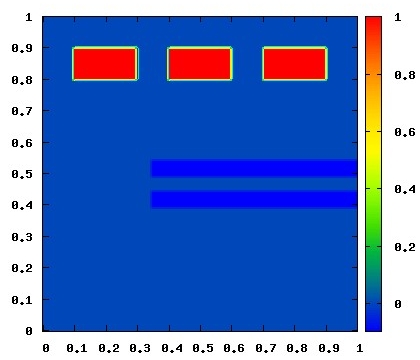} \\
	(i)
\end{minipage}
\begin{minipage}[c]{0.49\textwidth}
	\centering
	\includegraphics[width=0.49\textwidth,clip]{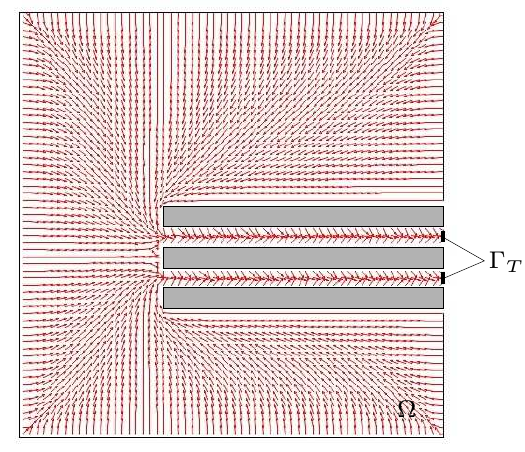}
	\includegraphics[width=0.49\textwidth,clip]{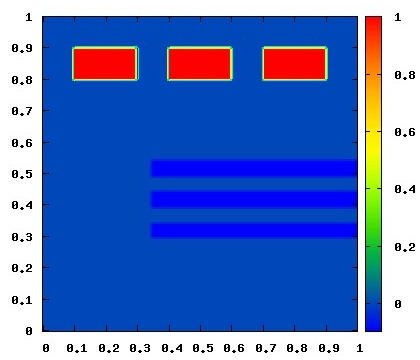} \\	
	(ii)
\end{minipage}
\caption{Desired velocity field and initial condition for the problem of the narrow passage (i) and of the two narrow passages (ii).}
\label{fig:narrowpass_desvel_initcond}
\end{figure}

\section{Application to some cases study}	\label{sect:appl}
In this section we apply the time-evolving measures model to some cases study, in order to test its ability in reproducing complex realistic features of two-dimensional pedestrian flows. To summarize, the model consists of Eq. \eqref{eq:pushfwd} set in $\Omega=[0,\,1]^2$, supplemented by an initial condition yielding the distribution $\mu_0$ of pedestrians at time $n=0$. The velocity field $v_n$ is of the form \eqref{eq:vn2}. In particular, we deduce the desired velocity $v_d$ from Eq. \eqref{eq:vd}, solving the Laplace equation \eqref{eq:Laplace} for the potential $u$ along with suitable boundary conditions that will be discussed from time to time, and the interaction velocity $\nu_n[\mu_n]$ from Eq. \eqref{eq:nun_imprvd} choosing $p_\nu[\hat{\mu}_n](x)=\beta\hat{\mu}_n(B_R^+(x))/R$.

In Subsects. \ref{subsect:desvel}, \ref{subsect:intvel} we have discussed the Lipschitz continuity of both $v_d$ and $\nu_n[\mu_n]$, which, together with the assumption $\mu_0\ll\Lebesgue^2$, ultimately enables us to apply Theorem \ref{theo:abs_cont} and speak of density of pedestrians $\rho_n\in L^1(\Omega)\cap L^\infty(\Omega)$ each $n>0$. Therefore, in the next applications we will constantly refer to these densities when showing the results of the numerical simulations.

\subsection{Pedestrian flow through a narrow passage}	\label{subsect:narrowpass}
In this first application we describe the motion of a crowd through a bottleneck formed by two long and thin obstacles, parallel and close to one another. The bottleneck may represent, for instance, a corridor or some stairs driving people toward an exit $\Gamma_T$ located at its end. The domain of the problem, along with the related desired velocity field of pedestrians, and the initial distribution of the crowd are shown in Fig. \ref{fig:narrowpass_desvel_initcond}(i). In particular, for the computation of $v_d$ we have set $u=0$ on all outer boundaries of the walking area and on the internal boundaries of the bottleneck, so as to get a repulsive effect that leads pedestrians to walk away from perimeter walls and to occupy the middle of the bottleneck. Conversely, we have prescribed $\nabla{u}\cdot\n=0$ on the remaining walls of the bottleneck to render a minor repulsion by the edges confining with its entrance, and $u=1$ on $\Gamma_T$ to identify the target of the crowd.

In view of the initial condition depicted in Fig. \ref{fig:narrowpass_desvel_initcond}(i), this problem can be compared with an analogous application at the microscopic scale proposed by Maury and Venel \cite{MaVe}, who model the flow of three groups of pedestrians directed toward a narrow escalator of a station after exiting a train.

\begin{figure}[t]
\centering
\begin{minipage}[c]{0.24\textwidth}
	\centering
	\includegraphics[width=\textwidth,clip]{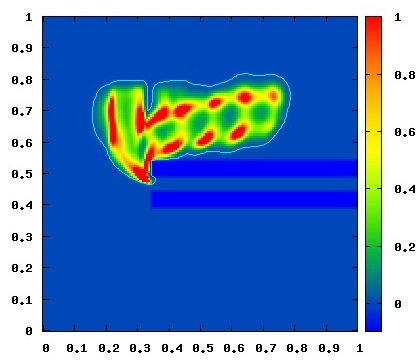} \\
	(i)
\end{minipage}
\begin{minipage}[c]{0.24\textwidth}
	\centering
	\includegraphics[width=\textwidth,clip]{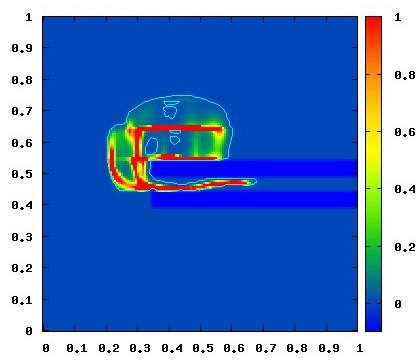} \\
	(ii)
\end{minipage}
\begin{minipage}[c]{0.24\textwidth}
	\centering
	\includegraphics[width=\textwidth,clip]{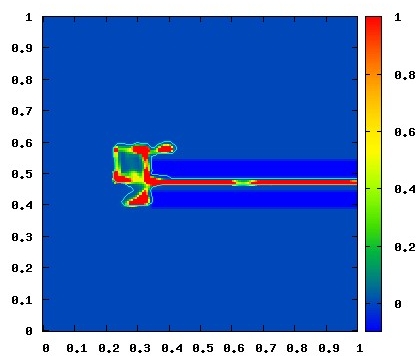} \\
	(iii)
\end{minipage}
\begin{minipage}[c]{0.24\textwidth}
	\centering
	\includegraphics[width=\textwidth,clip]{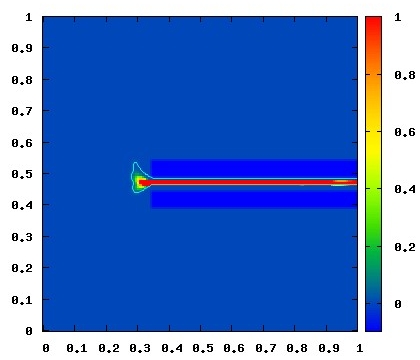} \\
	(iv)
\end{minipage}
\caption{Evolution of the density of pedestrians at successive time instants for the problem of the narrow passage.}
\label{fig:narrowpass_evol}
\end{figure}

Figure \ref{fig:narrowpass_evol}(i) shows that the three groups join in one single big group few instants after the beginning of the simulation, and then walk toward the entrance of the bottleneck from the left. Once there, they give rise to an obstruction (Fig. \ref{fig:narrowpass_evol}(ii)), with some people pushed also on the right side of the entrance (Fig. \ref{fig:narrowpass_evol}(iii)), until they all flow through the bottleneck (Fig. \ref{fig:narrowpass_evol}(iv)). Notice how the model predicts a self-organization of pedestrians in lanes of high density separated by areas of lower crowding. This is especially evident in Fig. \ref{fig:narrowpass_evol}(ii) for the group of pedestrians sliding along the upper edge of the first obstacle, and also in the bifurcation of the flow occurring in front of the entrance of the bottleneck.

\begin{figure}[t]
\centering
\begin{minipage}[c]{0.24\textwidth}
	\centering
	\includegraphics[width=\textwidth,clip]{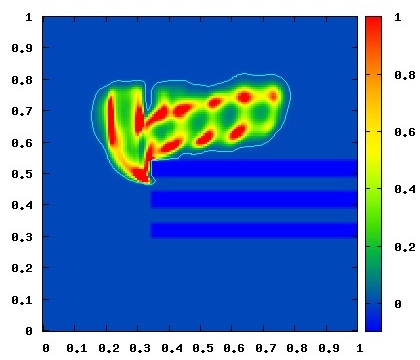} \\
	(i)
\end{minipage}
\begin{minipage}[c]{0.24\textwidth}
	\centering
	\includegraphics[width=\textwidth,clip]{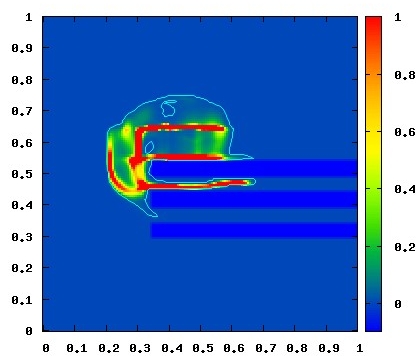} \\
	(ii)
\end{minipage}
\begin{minipage}[c]{0.24\textwidth}
	\centering
	\includegraphics[width=\textwidth,clip]{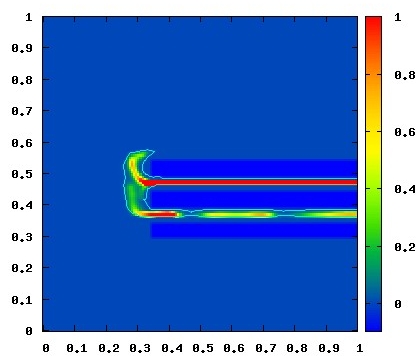} \\
	(iii)
\end{minipage}
\begin{minipage}[c]{0.24\textwidth}
	\centering
	\includegraphics[width=\textwidth,clip]{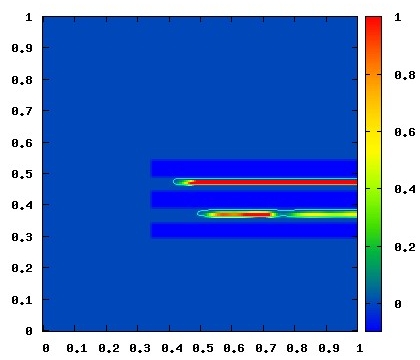} \\
	(iv)
\end{minipage}
\caption{Evolution of the density of pedestrians at successive time instants for the problem of two narrow passages.}
\label{fig:2narrowpass_evol}
\end{figure}

\subsection{Pedestrian flow through two adjacent narrow passages}	\label{subsect:2narrowpass}
We address now a case similar to that dealt with in Subsect. \ref{subsect:narrowpass}, but we add a second bottleneck beside the previous one, that equally leads to an exit (Fig. \ref{fig:narrowpass_desvel_initcond}(ii)), so that pedestrians have the choice of which passage to take in order to reach their target. Starting from the initial condition depicted in Fig. \ref{fig:narrowpass_desvel_initcond}(ii), the model predicts an evolution initially similar to that of the previous problem (compare Fig. \ref{fig:narrowpass_evol}(i) and Fig. \ref{fig:2narrowpass_evol}(i)). Later on, some individuals, pushed sideways by the crowd at the entrance of the first bottleneck, decide to take the other passage (Fig. \ref{fig:2narrowpass_evol}(ii)), which gives rise to a flow also in the second bottleneck. Pedestrians then start branching off in correspondence of the separation between the two bottlenecks (Fig. \ref{fig:2narrowpass_evol}(iii)), until all of them have entered either passage (Fig. \ref{fig:2narrowpass_evol}(iv)).

This problem may model, for instance, the flow of passengers exiting the cars of a subway and then heading for the exit of the metro station. Many stations have two (or even more) adjacent escalators that pedestrians can use to reach the ground level, however it is commonly observed that, on average, they prefer to take the closest one to their starting point. This behavior is clearly caught by the model, indeed Figs. \ref{fig:2narrowpass_evol}(ii-iv) show a lower crowd density in the farthest passage.

\begin{figure}[t]
\centering
\begin{minipage}[c]{0.24\textwidth}
	\centering
	\includegraphics[width=\textwidth,clip]{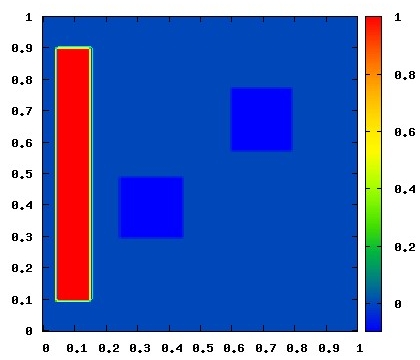} \\
	(i)
\end{minipage}
\begin{minipage}[c]{0.24\textwidth}
	\centering
	\includegraphics[width=\textwidth,clip]{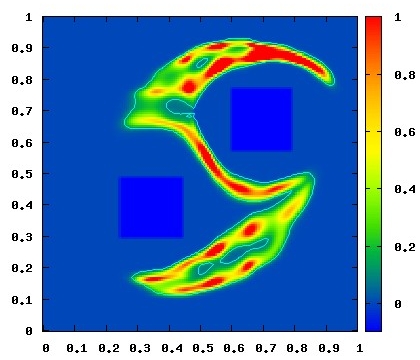} \\
	(ii)
\end{minipage}
\begin{minipage}[c]{0.24\textwidth}
	\centering
	\includegraphics[width=\textwidth,clip]{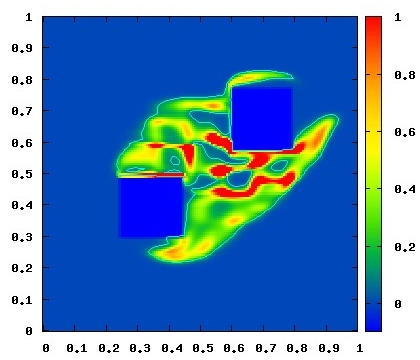} \\
	(iii)
\end{minipage}
\caption{Density of pedestrians in a domain with two obstacles, starting from the initial condition (i). The desired velocity field is that depicted in Fig. \ref{fig:bcu}, with (ii) Dirichlet and (iii) Neumann boundary conditions for the potential $u$ at the edges of the obstacles. Snapshots (ii) and (iii) refer to the same evolution time instant.}
\label{fig:effect_bc}
\end{figure}

\subsection{Effect of boundary conditions and modeling of obstacles}	\label{subsect:effect_bc}
Boundary conditions for the desired velocity field $v_d$ at the obstacle edges may sensibly affect the resulting configuration of the flow of pedestrians. Let us consider a walking area $\Omega$ featuring two obstacles $O_1,\,O_2$ and an exit $\Gamma_T$ located on a portion of the right vertical boundary, as illustrated in Fig. \ref{fig:bcu}. A compact group of pedestrians is initially positioned on the opposite side of the room (Fig. \ref{fig:effect_bc}(i)) and is guided toward the exit by either the field $v_d$ depicted in Fig. \ref{fig:bcu} left, with the potential $u$ set to zero at every internal and external boundary of $\Omega$ (but obviously $\Gamma_T$, where condition $u=1$ is prescribed instead), or the field $v_d$ depicted in Fig. \ref{fig:bcu} right, characterized by zero normal derivative of the potential $u$ at the obstacles. Figures \ref{fig:effect_bc}(ii), \ref{fig:effect_bc}(iii) show an instantaneous configuration of the distribution of pedestrians in either case at the same evolution time. The repulsive effect of the obstacles is particularly evident in the case of Dirichlet boundary conditions (Fig. \ref{fig:effect_bc}(ii)), with preferential paths bypassing both obstacles from the outside. Neumann conditions (Fig. \ref{fig:effect_bc}(iii)) produce instead the expected sliding of pedestrians along the edges, with a larger crowding of the area comprised between the two obstacles. This demonstrates that the set of boundary conditions to generate the desired velocity field has in general nontrivial consequences on the resulting flow of pedestrians, and may vary from case to case also in connection with the role played by each obstacle in every specific application.

\subsection{Lane formation versus clustering}	\label{subsect:lanes-clusters}
\begin{figure}[t]
\centering
\begin{minipage}[c]{0.24\textwidth}
	\centering
	\includegraphics[width=\textwidth,clip]{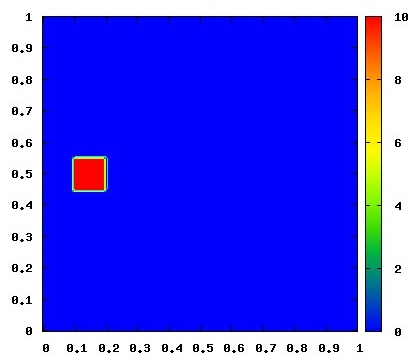} \\
	(i)
\end{minipage}
\begin{minipage}[c]{0.24\textwidth}
	\centering
	\includegraphics[width=\textwidth,clip]{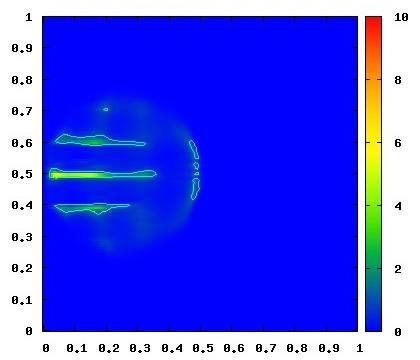} \\
	(ii-a)
\end{minipage}
\begin{minipage}[c]{0.24\textwidth}
	\centering
	\includegraphics[width=\textwidth,clip]{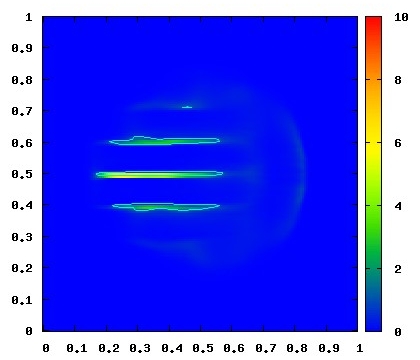} \\
	(iii-a)
\end{minipage} \\
\begin{minipage}[c]{0.24\textwidth}
	\centering
	\includegraphics[width=\textwidth,clip]{laneclust_00000.jpg} \\
	(i)
\end{minipage}
\begin{minipage}[c]{0.24\textwidth}
	\centering
	\includegraphics[width=\textwidth,clip]{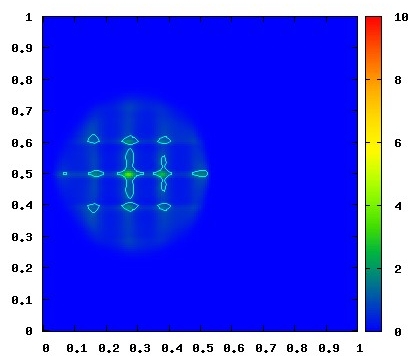} \\
	(ii-b)
\end{minipage}
\begin{minipage}[c]{0.24\textwidth}
	\centering
	\includegraphics[width=\textwidth,clip]{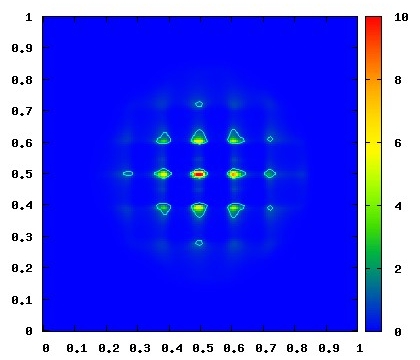} \\
	(iii-b)
\end{minipage}
\caption{Self-organization of a group of agents. Starting from the same initial condition (i), anisotropic interactions lead to lane formation (ii-a, iii-a), whereas isotropic interactions promotes clustering (ii-b, iii-b).}
\label{fig:laneclust_evol}
\end{figure}

We focus now on the self-organization properties predicted by our model for a system of agents characterized by either anisotropic or isotropic interaction mechanisms. Figure \ref{fig:laneclust_evol}(i) shows a compact group of agents initially located near the left side of the domain $\Omega$. We assume that they are driven by a constant rightward desired velocity field, which originates from the potential $u(x,\,y)=x$ solving Eq. \eqref{eq:Laplace} with the following set of boundary conditions:
\begin{equation*}
	\begin{cases}
		u=0 & \text{on\ } x=0 \\
		u=1 & \text{on\ } x=1 \\
		\nabla{u}\cdot\n=0 & \text{on\ } y=0,\,1.
	\end{cases}
\end{equation*}

Assume that the agents under consideration are pedestrians, who, as already mentioned, feature anisotropic interactions between each other for they are able to see ahead only. As shown in Fig. \ref{fig:laneclust_evol}(ii-a, iii-a), this fosters a self-organization of the mass in \emph{parallel lanes}, which follow the direction of the main velocity field and whose number depends on the initial distribution $\mu_0$ of pedestrians. We notice that the reciprocal distance between the lanes is comparable to the size of the neighborhood of interaction ($R=0.1$ in this simulation). Organization in uniformly walking lanes is actually observed in real pedestrian flows, as pointed out by Helbing and coworkers in \cite{HeFaMoVi,HeJo,HeMoFaBo}.

Let now the agents be able to see both ahead and behind, as it happens for instance with birds, whose view covers also a back area due to the lateral position of their eyes on the head. Interactions among the agents are then isotropic, as they do not depend on the desired direction of motion but are equally felt on the whole ball $B_R(x)$. In such a situation, starting from the same initial condition as before, our model predicts a \emph{clustering} of the mass (Fig. \ref{fig:laneclust_evol}(ii-b, iii-b)) while drifted by the main velocity field, with a characteristic distance between clusters again comparable to the size $R$ of the neighborhood of interaction. This kind of cooperative motion pattern is actually found in several multi-agent systems, such as birds or fishes, and in the specialized literature is commonly termed \emph{flocking} (cf. e.g., Krause and Ruxton \cite{KrRu} and the main references therein).

\subsection{Crossing flows}	\label{subsect:crossflow}
Finally, we study the problem of two groups of people walking in opposite directions toward one another. Several Authors have addressed this issue, especially from the microscopic point of view, finding in general a good agreement between the main features of this kind of flow and the predictions provided by their models. In the papers by Helbing and coworkers \cite{HeJo,HeMoFaBo} and by Hoogendoorn and coworkers \cite{HoDa1,HoDaBo}, which are a valuable source of empirical information on pedestrian behavior, particular attention is given to typical patterns emerging in crossing pedestrian flows. The microscopic models proposed by Helbing \emph{et al.} (\emph{ibidem}), Hoogendoorn and Bovy \cite{HoBo_ped}, Maury and Venel \cite{MaVe2} have proved to be successful in reproducing such experimental evidences.

In order to model crossing flows we need to introduce two measures $\mu_n^1,\,\mu_n^2:\B(\Omega)\to\R_+$ describing the space occupancy by either pedestrian group, and consequently two motion mappings $\gamma_n^1,\,\gamma_n^2:\Omega\to\Omega$ such that
\begin{equation}
	\mu_{n+1}^i(E)=\mu_n^i({(\gamma_n^i)}^{-1}(E)), \qquad \forall\,E\in\B(\Omega),\ i=1,\,2.
	\label{eq:pushfwd_2pop}
\end{equation}
Each group has its own desired velocity $v_d^i:\Omega\to\R^d$, which, consistently with the characterization stated in Subsect. \ref{subsect:desvel}, is unaffected by the desired velocity of the opposite group. Conversely, each interaction velocity depends now on both pedestrian distributions:
$$ \nu_n^i=\nu_n^i[\mu_n^1,\,\mu_n^2], \qquad i=1,\,2, $$
indeed the dynamics triggered by encounters among individual belonging to different populations plays a relevant role in this problem. Specifically, we assume that pedestrians of the first population feel uncomfortable when too close to pedestrians of the second population, due to their different walking targets, and vice versa. Therefore, they may decide to keep away from areas of high concentration of people coming in the opposite direction, trying instead to gain room for their walking direction. At the same time, people may still disagree with being overcompressed by pedestrians of the their own population, and therefore avoid an excessive proximity also with those walking in the same direction, or prefer instead to keep the contact with the latter in order to predominate over the oppositely walking population. To take these phenomena into account, we propose the following form of the interaction velocity:
\begin{equation}
	\nu_n^i[\mu_n^1,\,\mu_n^2](x)=\frac{1}{R}\lint_{B_R^+(x)}(x-y)\,d\sum_{j=1}^2\beta_{ij}\mu_n^j(y),
		\qquad i=1,\,2,
	\label{eq:nun-2pop}
\end{equation}
possibly along with an obvious extension of the correction \eqref{eq:nun_imprvd} at the boundaries of the domain, where:
\begin{itemize}
\item $\beta_{ij}\geq 0$ for $i\ne j$ determines the strength of the interaction of the population $i$ with the head-on population $j$;

\item $\beta_{ii}\in\R$ gives the strength of the interaction among people of the same population, with a disgregating effect if $\beta_{ii}>0$ and an aggregating effect if instead $\beta_{ii}<0$.
\end{itemize}

\begin{figure}[t]
\centering
\begin{minipage}[c]{0.24\textwidth}
	\centering
	\includegraphics[width=\textwidth,clip]{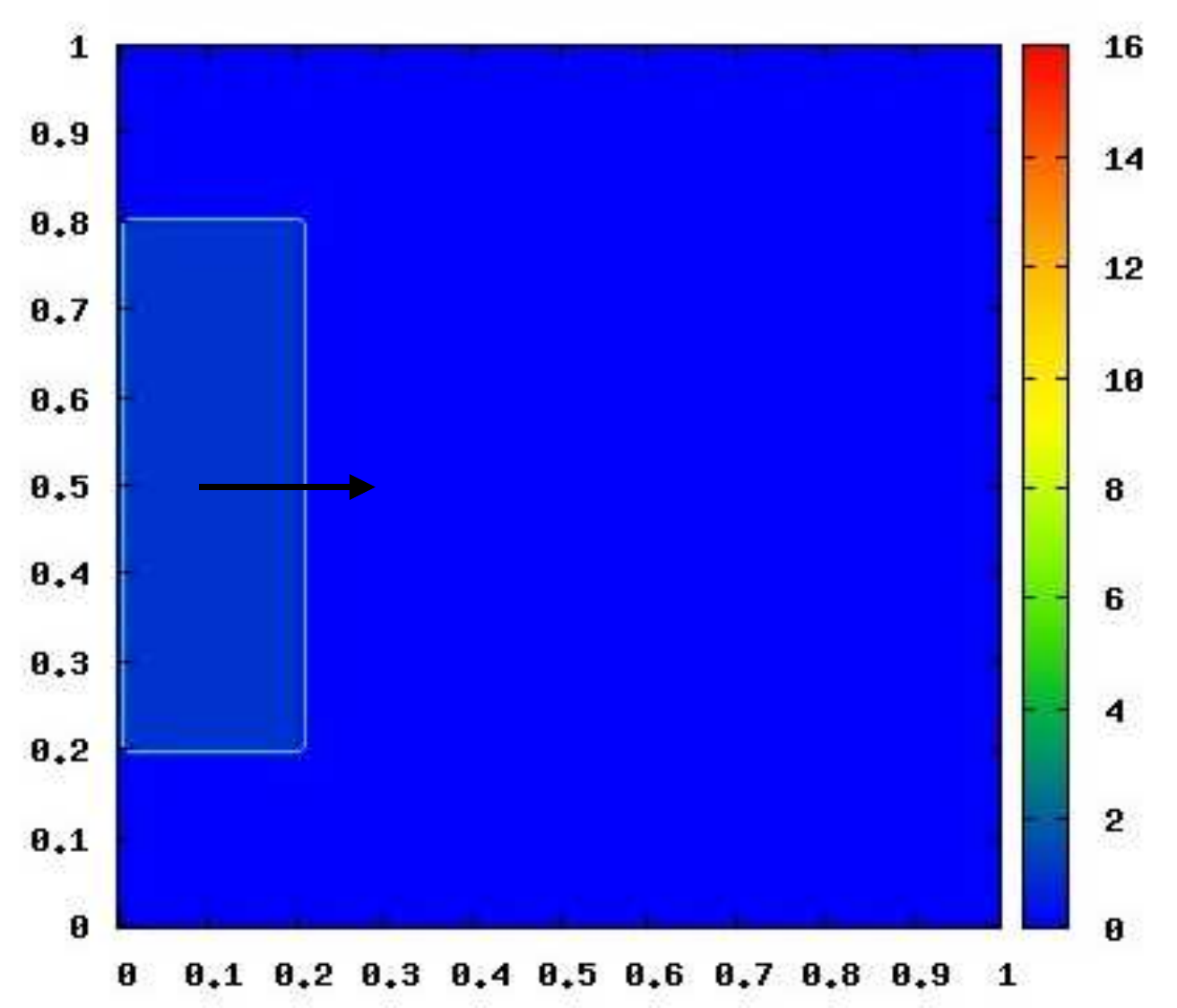}
\end{minipage}
\begin{minipage}[c]{0.24\textwidth}
	\centering
	\includegraphics[width=\textwidth,clip]{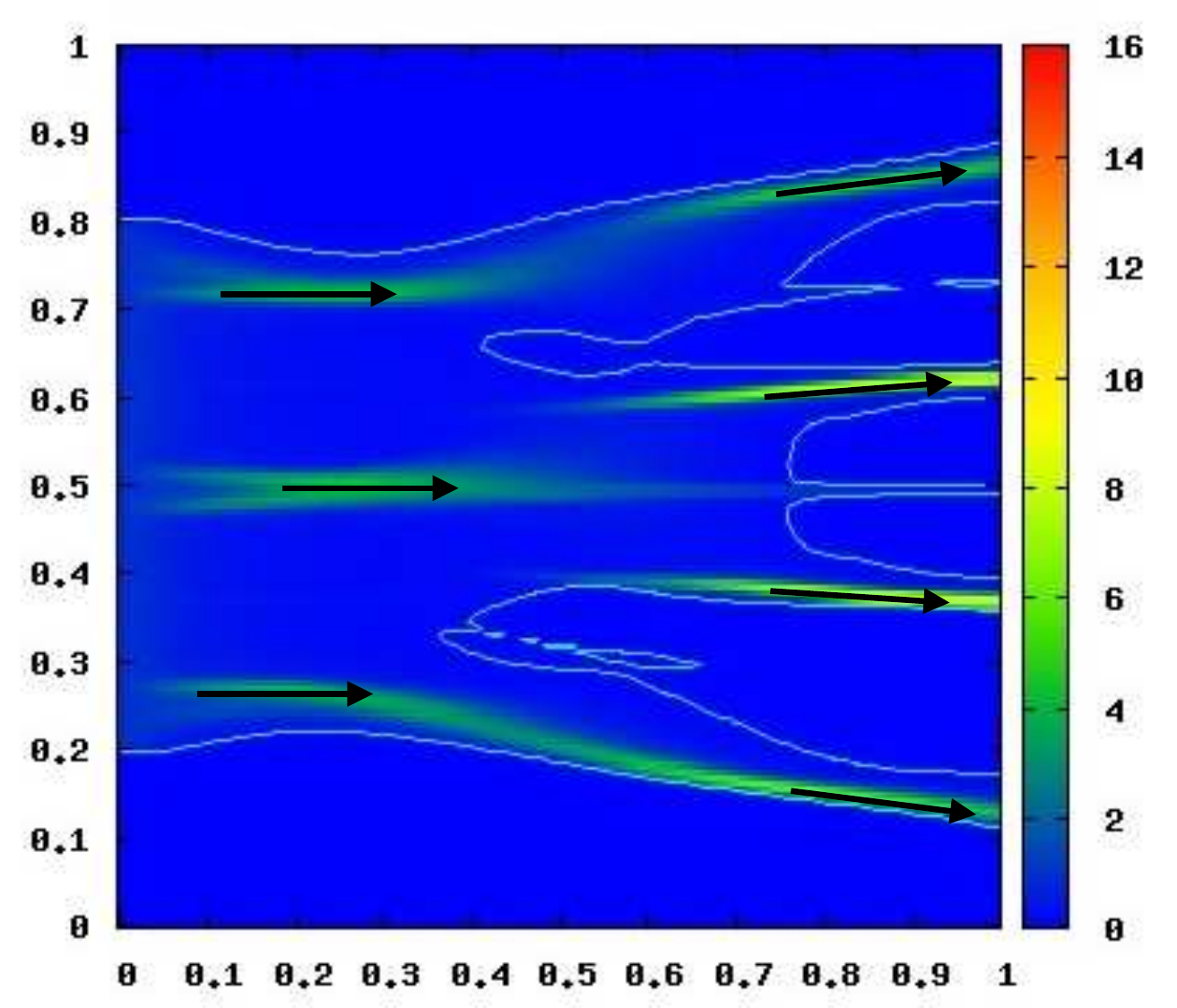}
\end{minipage}
\begin{minipage}[c]{0.24\textwidth}
	\centering
	\includegraphics[width=\textwidth,clip]{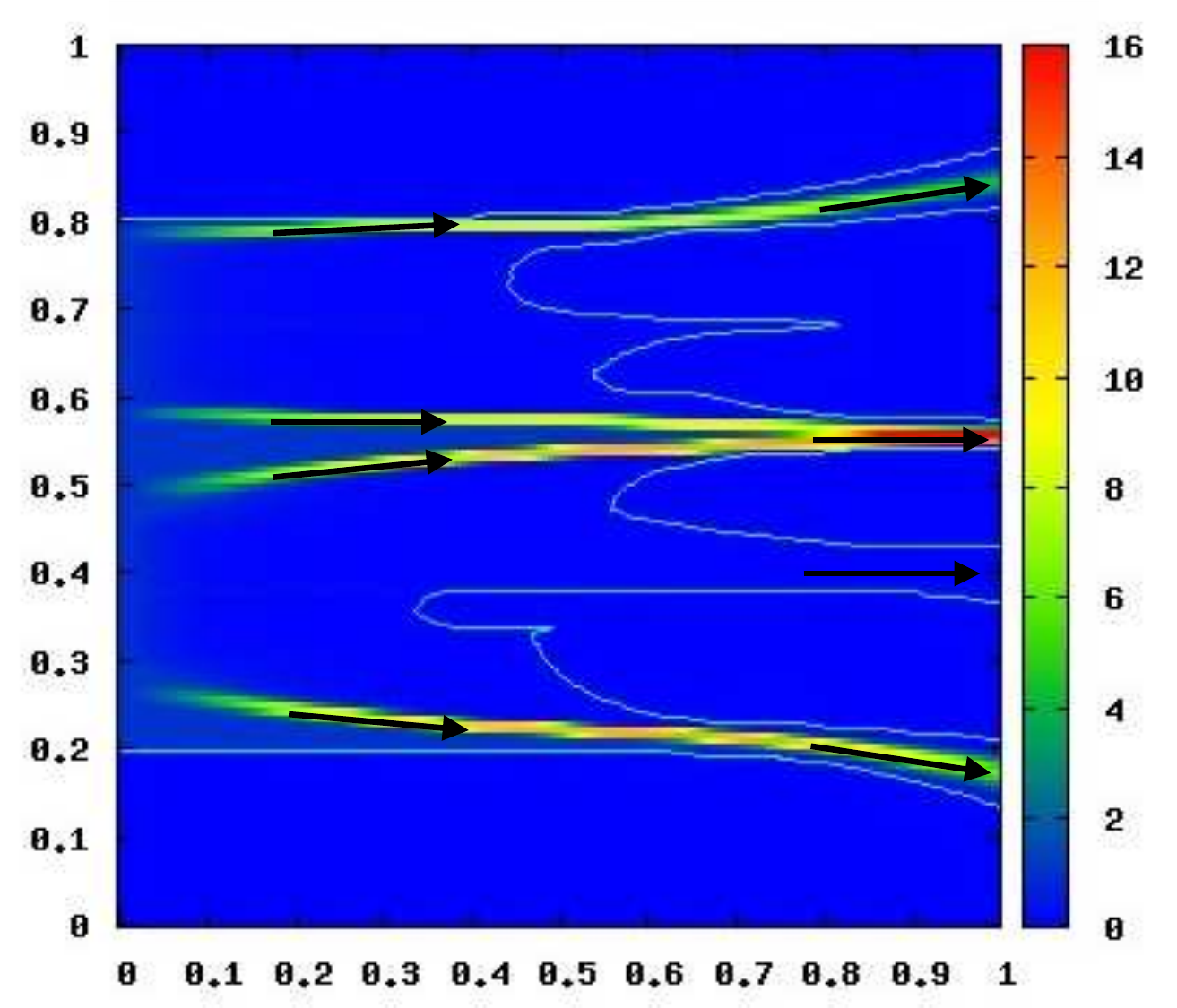}
\end{minipage} \\
\begin{minipage}[c]{0.24\textwidth}
	\centering
	\includegraphics[width=\textwidth,clip]{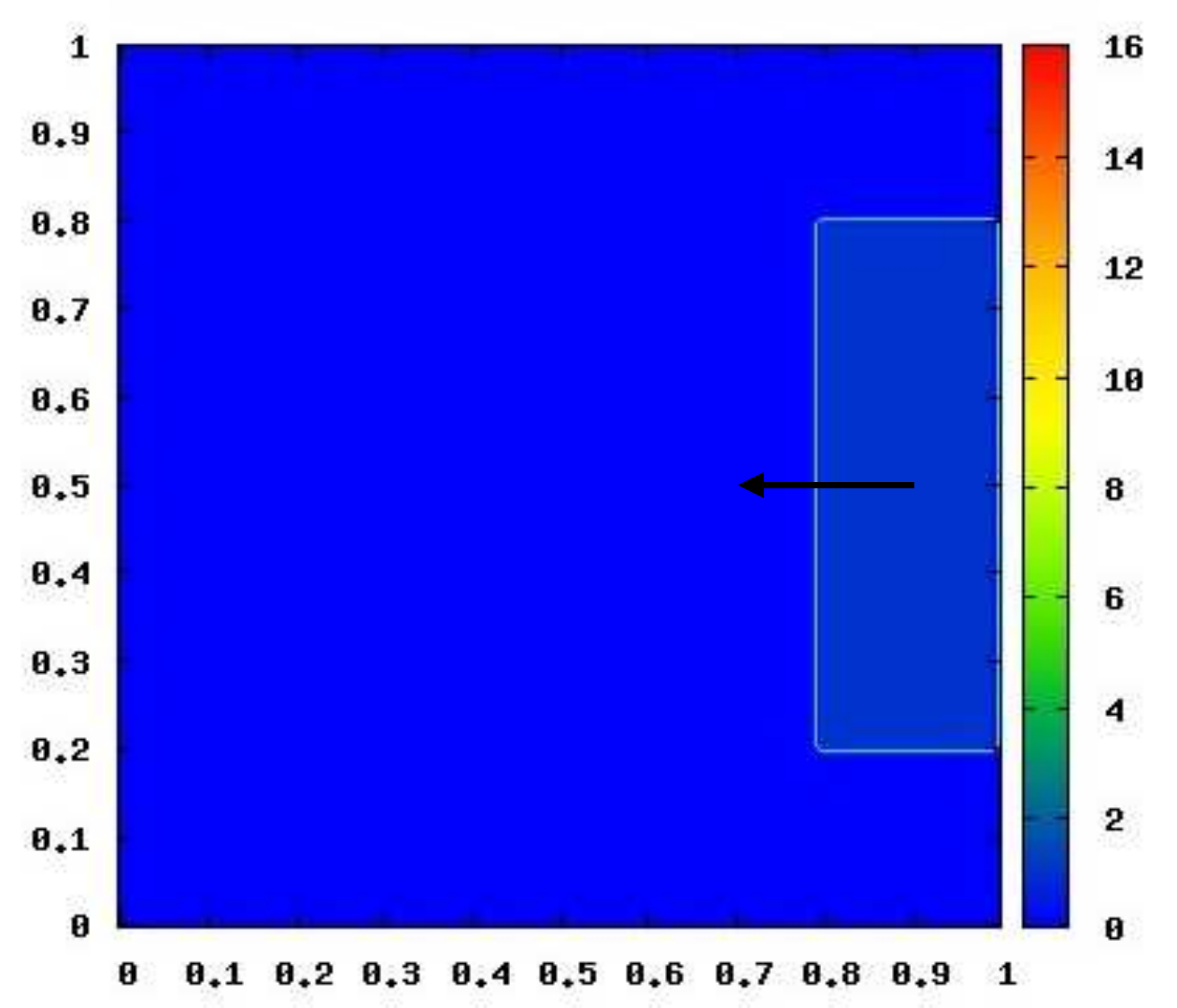} \\
	(i)
\end{minipage}
\begin{minipage}[c]{0.24\textwidth}
	\centering
	\includegraphics[width=\textwidth,clip]{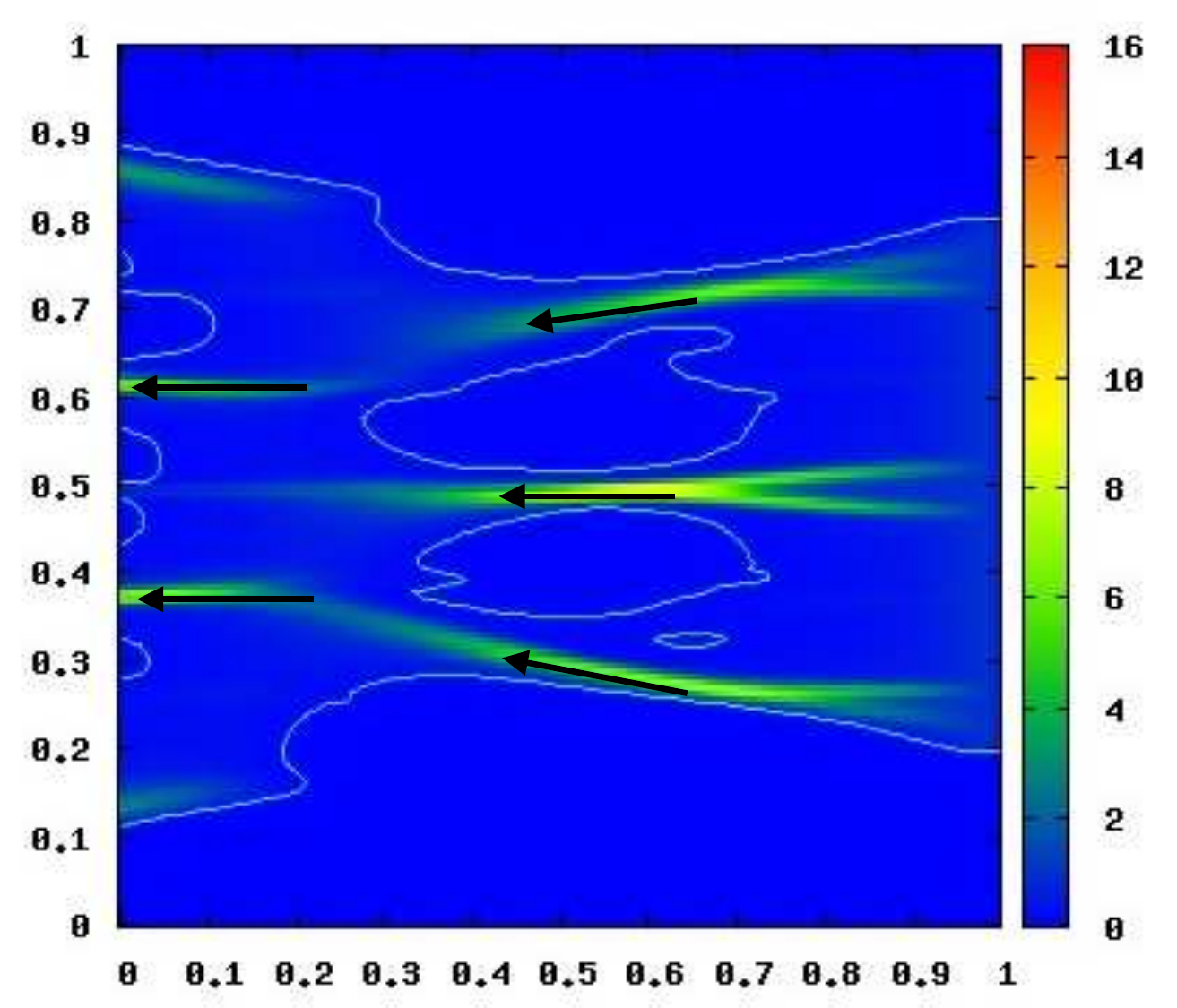} \\
	(ii)
\end{minipage}
\begin{minipage}[c]{0.24\textwidth}
	\centering
	\includegraphics[width=\textwidth,clip]{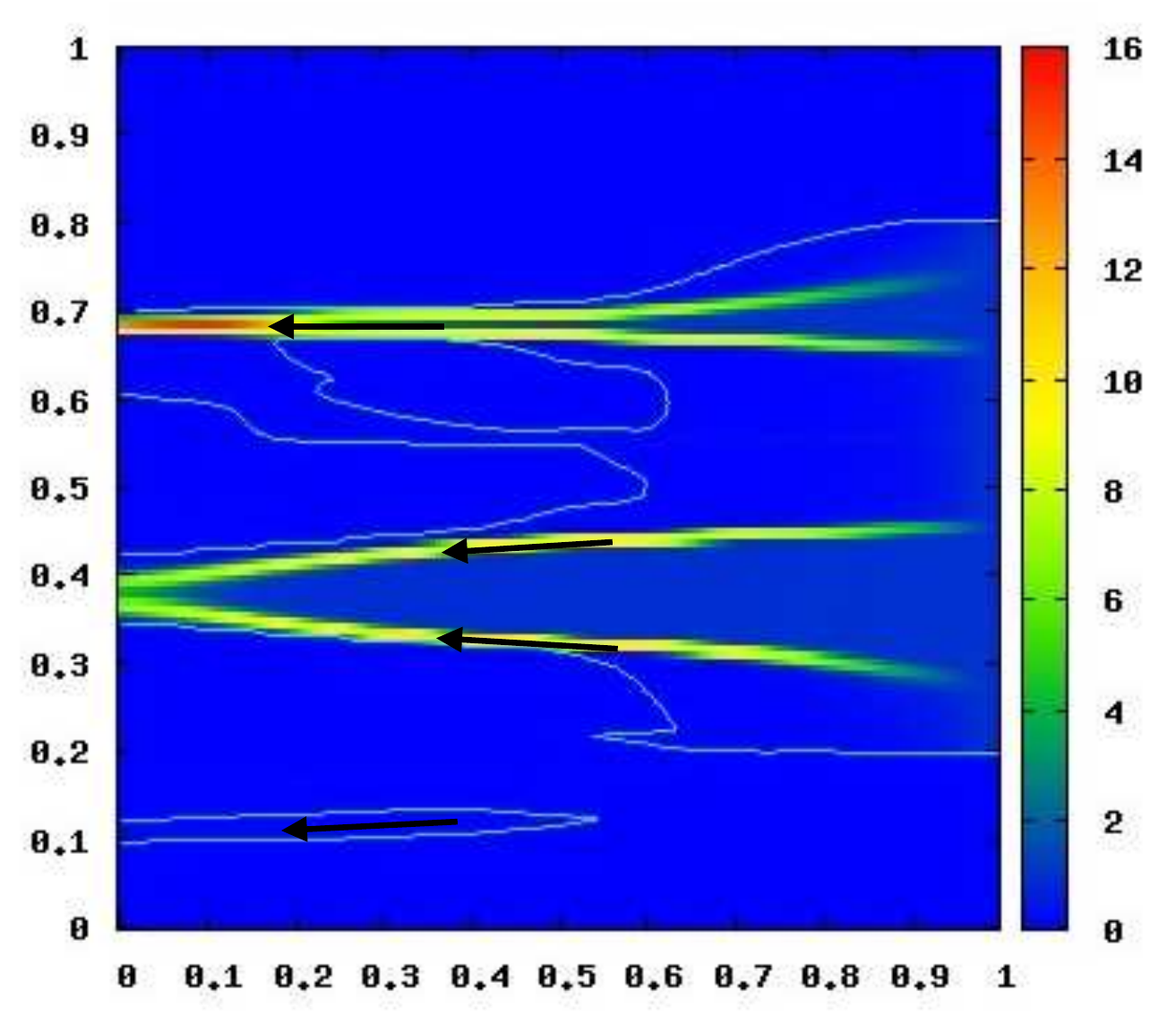} \\
	(iii)
\end{minipage}
\caption{(i) Two uninterrupted flows of pedestrians traveling rightward (upper row, population $i=1$) and leftward (lower row, population $i=2$) meet at the center of the domain. The two groups start to interact with each other, breaking the initial left-right symmetry (ii) and giving rise finally to a stable configuration of alternate uniformly walking lanes (iii). The cyan line in each plot is the $0.1$-level curve of the density $\rho_n^i$.}
\label{fig:oppflux}
\end{figure}

The mathematical problem generated by the two coupled equations \eqref{eq:pushfwd_2pop}, supplemented by proper initial conditions $\mu_0^1,\,\mu_0^2\ll\Lebesgue^d$, can be profitably addressed by making use of the techniques illustrated in \cite{PiTo} for nonlinear flows. In particular, the form \eqref{eq:nun-2pop} of the interaction velocity along with the regularity of the two (uncoupled) potential desired velocity fields $v_d^1$, $v_d^2$ give rise to motion mappings $\gamma_n^1$, $\gamma_n^2$ both complying with hypothesis \eqref{eq:prop_gamman} of Theorem \ref{theo:abs_cont}. This enables one to prove that two sequences of densities
$$ \{\rho_n^1\}_{n>0},\, \{\rho_n^2\}_{n>0}\subseteq L^1(\Omega)\cap L^\infty(\Omega) $$
exist, each of which is unique, such that $\rho_n^i\geq 0$ a.e. in $\Omega$ and $d\mu_n^i=\rho_n^i\,dx$ each $i=1,\,2$ and each $n>0$.

Figure \ref{fig:oppflux} shows the prediction of the model for two uninterrupted flows of pedestrians entering the domain from the left and the right boundary, respectively. In this simulation, we have set $\beta_{11}=\beta_{22}=0$ (no interaction among pedestrians belonging to the same population), so as to focus specifically on the effect of the interactions between the two oppositely walking groups. The results clearly demonstrate the ability of the model to reproduce first the rupture of the left-right symmetry, as soon as the two groups start to interact, and then the emergence of typical patterns of alternate uniformly walking lanes, which are experimentally observed as a characteristic self-organization phenomenon in crossing flows and, at present, mathematically described mainly by means of microscopic models (cf. the above-cited papers).

It is worth noticing that our mathematical model has \emph{not} been conceived with the specific purpose of describing either the self-organization in lanes/clusters discussed in Subsect. \ref{subsect:lanes-clusters} or the emerging uniformly walking lanes addressed here. Rather, we stress that the model is able to catch all these behavioral customs as a by-product of much more general (and even more elementary) modeling principles, resorting ultimately to the basic idea of nonlocal interactions among pedestrians.

\section{Conclusions and research perspectives}	\label{sect:conclusions}
In this paper we have systematically applied the mathematical structures developed in \cite{PiTo} to the modeling of pedestrian flows. The reference framework consists of a discrete-time sequence of Radon positive measures $\{\mu_n\}_{n\geq 0}$ defined over the Borel $\sigma$-algebra $\B(\Omega)$, $\Omega\subset\R^d$ being the domain of the problem, which evolve according to the recursive relation $\mu_{n+1}=\gamma_n\#\mu_n$ (\emph{push-forward}). The core of the structure is represented by the \emph{motion mappings} $\gamma_n:\Omega\to\Omega$, $\gamma_n(x)=x+v_n(x)\Delta{t}$, which describe the dynamics of the system. Here, $v_n:\Omega\to\R^d$ is the velocity field of pedestrians and $\Delta{t}>0$ the time step. The regularity of the measures $\mu_n$, specifically the fact of being absolutely continuous with respect to the Lebesgue measure $\Lebesgue^d$ on $\R^d$, depends on some non-singularity properties of the $\gamma_n$'s, which can be rephrased in that they do not cluster the Lebesgue measure, i.e., do not map Lebesgue non-negligible sets into Lebesgue negligible sets. This guarantees the existence and uniqueness of densities $\{\rho_n\}_{n>0}\subset L^1(\Omega)$ such that $d\mu_n=\rho_n\,dx$, which are also bounded in $\Omega$, i.e., $\rho_n\in L^\infty(\Omega)$ each $n>0$, provided $\rho_0$ is (cf. Theorem \ref{theo:abs_cont}). The existence of densities is particularly significant for application purposes, as it allows to establish a direct connection of the measure-theoretical structures with classical methods of continuum mechanics. Indeed, not only can the relation $\mu_{n+1}=\gamma_n\#\mu_n$ be equivalently rewritten in a form which reminds of the usual structure of conservation laws (cf. Eq.\eqref{eq:conslaw}), but, considering that the push forward of the $\mu_n$'s turns out to be the direct time discretization of the abstract continuous-time conservation law (cf. \cite{PiTo})
\begin{equation*}
	\frac{d}{dt}\mu_t(\gamma_t(E))=0, \qquad \forall\,E\in\B(\Omega),
\end{equation*}
one is led, under the \emph{continuum hypothesis} $\mu_t,\,\mu_n\ll\Lebesgue^d$ each $t>0$, $n>0$, to a time discretization of the equation
\begin{equation*}
	\frac{d}{dt}\lint_{\gamma_t(E)}\rho(t,\,x)\,dx=0, \qquad \forall\,E\in\B(\Omega),
\end{equation*}
i.e., the classical (Lagrangian) form of the mass conservation equation dealt with by continuum mechanics.

Numerical approximation of the push forward can be performed by a suitably devised scheme (cf. Eq. \eqref{eq:numscheme}), introducing piecewise constant measures $\lambda_h^n:\B(\Omega)\to\R_+$ over a partition of the domain $\Omega$ with characteristic size individuated by a parameter $h>0$. Under some technical assumptions, one can control the so-called \emph{localization error} (cf. Subsect. \ref{subsect:numerics}) produced by the $\lambda_h^n$'s over the $\mu_n$'s in terms of the grid size $h$ (cf. Theorem \ref{theo:stability}), provided the latter is chosen with respect to the time step $\Delta{t}$ in such a way to fulfill a CFL-like condition (cf. Eq. \eqref{eq:CFL}).

It is worth stressing that this theoretical setting enables one to address both analytical and numerical issues in a thorough and unified manner for whatever dimension $d\geq 1$ of the spatial domain. The same is not true for classical approaches based on nonlinear hyperbolic conservation laws, which are well known to generate nontrivial analytical and numerical difficulties as soon as the spatial dimension of the problem gets greater than $1$. This is particularly relevant for the application to pedestrian flows, which develop mainly in two dimensions: The time-evolving measures framework allows a qualitative mathematical mastery of the problem that nonlinear conservation laws might hardly achieve. As an example of the difficulties raised by standard modeling techniques, let us consider the handling of boundary conditions. From the modeling point of view, one has to guarantee that pedestrians do not either flow through obstacles or leave the walking area from portions of the boundary other than the prescribed outlet regions. On the other hand, when using hyperbolic partial differential equations, the definition of the outflow portion of the boundary is not simply a matter of appropriate boundary conditions, because at each time instant one must also take into account the orientation of the characteristic velocities of the problem. Therefore, it may be argued that some unilateral constraints on the velocity should be imposed at the boundary, that the solution of the problem must then comply with, for instance $v\cdot\n\lesseqgtr 0$ on some portions of $\partial\Omega$, in order to ``deflect'' characteristics where needed. Clearly, this introduces additional technicalities, that can be instead ruled out elegantly by a time-evolving measures approach, as indicated in Subsect. \ref{subsect:intvel}.

The main modeling task posed by the time-evolving measures setting \eqref{eq:motmap}-\eqref{eq:pushfwd} is the definition of the velocity field $v_n$. For pedestrian flows, inspired by some analogous considerations proposed by Maury and Venel \cite{MaVe}, we have suggested a structure accounting for two main factors affecting the motion of walkers, namely:
\begin{itemize}
\item Pedestrians' will to reach a specific target placed in the walking area, expressed by a \emph{desired velocity field} $v_d:\Omega\to\R^d$, which essentially depends on the geometry of the domain, including the presence of intermediate obstacles to be bypassed. This is the velocity a person would have in the absence of other people in the surroundings, and is modeled as a conservative vector field coming from a scalar potential.
\item Pedestrians' tendency to avoid congested areas, expressed by an \emph{interaction velocity} $\nu_n[\mu_n]:\Omega\to\R^d$, which depends in a nonlocal way on the distribution $\mu_n$ of pedestrians themselves. To model this term, we have borrowed some ideas from the Eulerian approach to the analysis of \emph{rendez-vous} problems for multi-agent systems by Canuto \emph{et al.} \cite{CaFaTi}.
\end{itemize}
The superposition of these two effects, plus some suitable corrections at the boundaries of $\Omega$ and of the obstacles illustrated in Subsect. \ref{subsect:intvel}, yields the final form of pedestrians' velocity:
$$ v_n[\mu_n](x)=v_d(x)+\nu_n[\mu_n](x), \qquad x\in\Omega, $$
which closes the model.

By means of these structures, we have addressed some representative cases study with the aim of testing the ability of the model to reproduce complex features of pedestrian flows pointed out in the specialized experimental literature (see e.g., Helbing \emph{et al.} \cite{HeFaMoVi,HeJo,HeMoFaBo}, Hoogendoorn \emph{et al.} \cite{HoDa1,HoDaBo}, Krause and Ruxton \cite{KrRu}, and their main references). In more detail, we have considered several applications to motion in domains with obstacles, possibly also with more than one target for pedestrians, mimicking realistic situations like the access to one or two escalators by a group of people getting off the cars of a subway (cf. Subsects. \ref{subsect:narrowpass}, \ref{subsect:2narrowpass}) or the outflow of a crowd from a room with two pillars partially hiding the exit (cf. Subsect. \ref{subsect:effect_bc}). In the latter case, we have also studied the effect of different boundary conditions for the desired velocity field at the edges of the obstacles on the flow. In addition, we have shown that the model is able to reproduce interesting self-organization phenomena of human crowds, like the formation of dynamic lanes in a group of pedestrians sharing the same desired direction of motion (cf. Subsect. \ref{subsect:lanes-clusters}) and the emergence of alternate uniformly walking lanes in crossing flows (cf. Subsect. \ref{subsect:crossflow}).

It may be questioned that, unlike standard macroscopic models of continuum mechanics, for instance those using the formalism of partial differential equations, the modeling structures by time-evolving measures presented in \cite{PiTo} and in the present paper do not allow to deal with continuous-time dynamics. In other words, we are currently lacking a ``limit model'' for $\Delta{t}\to 0$. It is worth pointing out that, as far as applications are concerned, this is actually not a dramatic drawback, because the framework we have introduced still provides conceptual tools for a mathematical study of pedestrian flow problems, which have proved to be successful in addressing numerous applications to real cases. On the other hand, the great reward for the price paid of discrete time modeling is the possibility to approach in a unified manner, from both the analytical and the numerical point of view, $d$-dimensional systems with no additional technical difficulties when $d>1$. We stress the importance of this both to attain a satisfactory theoretical mastery of the modeling structures and to treat immediately realistic two-dimensional applications, without the need for devising preliminarily simplified explorative one-dimensional approximations. In general, the same is not as straightforward for more classical modeling techniques based on nonlinear conservation laws. Finally, we observe that the investigation of the limit behavior of the model for $\Delta{t}\to 0$ is, however, a possible by-product of this research line, which has then also the merit of promoting new ways of theoretical speculation.

Besides theoretical issues, further research developments may involve the use of the model of pedestrian flows presented here in connection with control and optimization topics. It is known that obstacles, which in the present work have been regarded as essentially passive elements of the walking area, often play instead an active role in forcing the direction of motion of the walkers. As an example, they may be designed and positioned so as to foster the flow of the crowd along specific directions, as it happens for instance in shopping centers when buyers are induced to move close to particularly attractive zones, or to improve the safety of pedestrians when accessing/leaving certain areas. In this context, we would like to recall \emph{Braess' paradox} for pedestrian flows, which has been pointed out by Hughes \cite{Hu2} and then drawn by several other Authors. Such a paradox, credited to the mathematician Dietrich Braess, was originally formulated for traffic flows on networks: It states that adding extra capacity (for instance, a further edge) to a network can in some cases reduce the overall performance (see e.g., \cite{Wiki:Braess} for more details). Rephrased in the abstract, it means that a condition intuitively expected to lead to an improvement may instead give rise to worse outcomes. Hughes suggests to invert Braess' paradox for pedestrian flows, after observing that placing an obstacle in front of an exit (intuitively worse condition) may sometimes improve the stream of people from a room (better outcome). The reason for this unexpected fact might be due to a sort of reorganization of the flow of pedestrians induced by the obstacle, provided  shape, size, and positioning of the latter are accurately studied. Clearly, the mathematical treatment of such issues needs to take advantage of sufficiently handy, but also realistic, two-dimensional models of pedestrian flows. In this respect, we believe that our macroscopic model by time-evolving measures can be profitably used to ground specific control and optimization problems of practical interest for real word applications.

Other research perspectives can be developed on the modeling side. As reviewed in Sect. \ref{sect:soa}, some macroscopic models already available in the literature (cf. Bellomo and Dogb\'e \cite{MR2438218} and the main references therein) describe the flow of pedestrians by both the conservation of mass and the balance of linear momentum, using Navier-Stokes-like partial differential equations with nonclassical force contributions. A natural question is whether one can do the same in a measure-theoretical setting, specifically what could be a measure-theoretical counterpart of the balance law
$$ \partial_t(\rho v)+\nabla\cdot(\rho v\otimes v)=F[\rho,\,v], $$
which, as it is well-known in continuum mechanics, is equivalent to the second equation of system \eqref{eq:BellDog} under the assumption of conservation of the mass. The question gets even more interesting if one considers that the nonlinearity of the convective term $\nabla\cdot(\rho v\otimes v)$ makes Navier-Stokes equations able to catch the possible appearance of turbulent regimes in the fluid, which Helbing \emph{et al.} \cite{HeJoAA} pointed out to play a role also in the flow of pedestrians. It may be argued that a careful translation of the continuum linear momentum balance equation in the language of time-evolving measures could enable our model to describe also turbulence phenomena in the motion of human crowds. The transition from laminar to turbulent flow could then be interpreted as the ability of the model to reproduce the emergence of panic.

\bibliographystyle{plain}
\bibliography{/dati/Ricerca/Bibliografie/Pedestrian/pedestrian-biblio,/dati/Ricerca/Bibliografie/Traffico/traffic-biblio}
\end{document}